\documentclass[12pt,a4paper]{article}
\usepackage{fleqn,amsfonts,amssymb,amsmath,amscd,epsfig,version,theorem} %
\usepackage{fancyhdr,enumerate,epsfig}
\usepackage{yhmath}
\usepackage{hyperref}
\usepackage[commandnameprefix=always]{changes}
\definechangesauthor[name=<AK>]{01}
\definechangesauthor[name=<MQ>]{02}
\numberwithin{equation}{section}
\numberwithin{figure}{section}
\theoremstyle{change}
\newtheorem{theorem}{Theorem} [section]
\newtheorem {lemma}[theorem]{Lemma}
\newtheorem {proposition}[theorem]{Proposition}
\newtheorem {corollary}[theorem]{Corollary}
\newtheorem {defi}[theorem]{Definition}
{\theorembodyfont{\normalfont}
{\theorembodyfont{\normalfont}\newtheorem {example}[theorem]{Example}
{\theorembodyfont{\normalfont}
{\theorembodyfont{\normalfont}\newtheorem {remark}[theorem]{Remark}
{\theorembodyfont{\normalfont}\newtheorem {remarks}[theorem]{Remarks}
\newcommand{\beq}{\begin{equation}}
\newcommand{\eeq}{\end{equation}}
\newcommand{\Leq}[1]{\label{#1}\end{equation}}
\newcommand{\beqn}{\begin{eqnarray}}
\newcommand{\eeqn}{\end{eqnarray}}
\newcommand{\beqno}{\begin{eqnarray*}}
\newcommand{\eeqno}{\end{eqnarray*}}
\newcommand{\es}{\emptyset}
\renewcommand {\l}{\left}
\newcommand {\ri}{\right}
\newcommand {\eps}{\epsilon} 
     
\newcommand {\LA}{\left\langle}
\newcommand {\RA}{\right\rangle}
\newcommand {\eh}{{\textstyle \frac{1}{2}}}

\newcommand {\bN}{{\mathbb N}}
\newcommand {\bR}{{\mathbb R}}

\newcommand {\bQ}{{\mathbb Q}}
\newcommand {\bS}{{\mathbb S}}

\newcommand{\idty}{{\rm 1\mskip-4mu l}} 
\newcommand{\cC}{{\cal C}}  
\newcommand{\cD}{{\cal D}} 
\newcommand{\cE}{{\cal E}} 
\newcommand{\cM}{{\cal M}}
\newcommand{\cP}{{\cal P}}
  
\newcommand{\cT}{{\cal T}}
\newcommand{\bem}{\l(\! \begin{array}}
\newcommand{\eem}{\end{array}\!\ri)}
\newcommand{\bsm}{\left(\begin{smallmatrix}} 
\newcommand{\esm}{\end{smallmatrix}\right)}  
\newcommand{\NN}{\nonumber}

\newcommand{\qmbox}[1]{\quad\mbox{#1}\quad}
\renewcommand {\max}{{{\rm max}}}
\newcommand{\hSE}{{\widehat{\Sigma}_E}}

\newcommand{\Di}{\Delta^I} 
\newcommand{\CS}{\mathrm{CS}}
\newcommand{\norm}[1]{\left\| #1 \right\|}
\usepackage{color}

\begin{document}
\title{Nondeterministic particle systems}
\author{Andreas Knauf\thanks{Department Mathematik, 
Friedrich-Alexander-Universit\"at Erlangen-N\"{u}rnberg,\newline
\hspace*{4mm} Cauerstr.\ 11, D--91058 Erlangen, Germany. } 
\and 
Manuel Quaschner\thanks{Institut f\"ur Mathematik, Friedrich Schiller Universit\"at Jena,
07737 Jena, Germany.\newline \hspace*{4mm} \href{mailto:knauf@math.fau.de} {\tt \{knauf,quaschner\}@math.fau.de}}} 
\date{\today}
\maketitle
\vspace*{-10mm}
\begin{abstract}
We consider systems of $n$ particles that move with constant velocity between collisions.
Their total momentum but not necessarily their kinetic energy is preserved at collisions.
As there are no further constraints, these systems are nondeterministic. 
In particular we examine trajectories with infinitely many collisions.
\end{abstract}
\tableofcontents
%
\section{Introduction} \label{sec:introduction}
%
{\bf Motivation:} 
The system described in the {\em abstract} 
models the kinematic question about the consequences of
conservation of total momentum in multi-particle systems. 
If one would additionally demand conservation of energy, then
it would exhibit an upper bound on the number of collisions, only depending on the  
masses of the $n$ particles \cite{FKM}. Such an upper bound does not exist 
here (if $n \ge 3$).\\
Conservation of both total momentum {\em and} energy are elementary 
properties of particle systems. 
So, before presenting our results, the question arises whether
our model does nevertheless describe traits of realistic $n$--particle systems,
which interact deterministically via pair potentials. This is indeed the case, 
if some of the particles are tightly bound and if one only observes the momenta
of these clusters, and not their internal energy.
\begin{example}[Celestial mechanics]\quad\\ 
An example is provided by the $n$--body problem of celestial mechanics.
There, of course, total momentum {\em and} total energy are preserved
by the dynamics.
\begin{enumerate}[$\bullet$]
\item 
$n=2$ particles with positive total energy move on Kepler hyperbolae that
are asymptotic to straight lines. Yet, after a near-collision, 
their {\em directions} are hard to predict if one does not know
the initial conditions precisely.
\item 
For $n=3$ consider 
clusters composed of a single body and a binary, respectively, 
also experiencing a near-collision. 
In the future, the clusters could separate again and move 
asymptotically on near straight lines. But without precise knowledge of the initial
conditions one cannot even predict their {\em speed}. This is because 
during near-collision, the internal energy of the nontrivial cluster 
can change by an arbitrary amount, which
equals the change of kinetic energy, up to sign, see {\sc McGehee} \cite{McG}.
\hfill $\Diamond$
\end{enumerate}
\end{example}
So our nondeterministic billiard can describe some traits of $n$--body 
problems. In fact, our main motivation was to model {\em non-collision singularities}
in celestial mechanics, where particles move to spatial infinity in finite time.
These are known to exist for $n\ge 4$ bodies, see {\sc Xia}  \cite{Xia} and  
{\sc Xue} \cite{Xue}, and the
solutions of the non-deterministic model with infinitely many collisions
could serve as a kind of skeleton.
\bigskip\\
{\bf Contents:}
We set the stage in Section \ref{sec:notation} 
where we fix notation, mainly concerning the
combinatorics and kinematics of $n$ particles.\\
Their nondeterministic dynamics is formally introduced in Definition \ref{def:ND}:
For the center of mass configuration space $M$ of $n$ particles in $d$ dimensions, 
a {\em nondeterministic trajectory} $q \in C(I, M)$ is defined on a maximal interval
$I=(T^-,T^+)$. 

On the set ${\cal T}\subseteq I$ of {\em collision times}, the clustering changes.
At times $t\in I$ the {\em collision set} $\mathrm{CS}(t)$ is the set partition of 
$\{1, \ldots, n\}$ that corresponds to the clustering.
The momentum $p \in M^*$ is constant on the components of $I\backslash {\cal T}$,
and the cluster barycenters for $\mathrm{CS}(t)$
have constant momenta near points $t \in {\cal T}$.

As kinetic energy is not preserved, any projection $\bR^d\to \bR^d$ 
transforms a nondeterministic trajectory
into a (part of) a nondeterministic trajectory, too 
(Lemma \ref{lem:projection} and Remark \ref{rem:projection}).

An important physical quantity that we study is the total moment of inertia 
$J=\eh \|q\|^2_{\mathcal{M}}$ of the particle system (Theorem \ref{thm:finer}), 
It is convex in its time dependence, which has several 
implications. In particular the moments where there is a change of clustering
form a nowhere dense countable set (Remark \ref{rem:J:T}). 
$T_{\mathrm{coll}} \in {\cal T}$ is called a {\em total collision time} 
if $q(T_{\mathrm{coll}}) = 0$.
By convexity there can be at most two such events 
(the second one is then an ejection). \\
Already for three particles, ${\cal T}$ can have accumulation points at  
{\em escape times} $T^\pm$ and at total collision times 
(Remark \ref{rem:accumulation}).  
For $n \ge 4$ it may have a larger set of accumulation points, but these are 
organised hierarchically by the partition poset of the particles.

Three nondeterministic particles exhibit a relatively simple behaviour.
If their moment of inertia grows initially, it does so exponentially in the number
of collisions (Example \ref{ex:3:particles}). This is not anymore the case if $n\ge4$, as
then subsystems of an expanding system can contract. 
The times where all particles interacted with
one another -- if only indirectly -- are called {\em chain-closing}. One major result
(Theorem \ref{thm:ExpLowerBound}) is that, under some hypotheses, size and 
speed of the expansion is exponential
in the number of chain-closing events.  

In Proposition \ref{prop:conv:trajectories}
we show that, similarly to the deterministic case, locally uniform limits of 
sequences of nondeterministic trajectories are trajectories, too.\\
Finally, we treat Lagrangian relations for multiple scattering events in 
Section~\ref{sec:Lagrangian}. 
That is, instead of considering individual nondeterministic trajectories, we embed them into families. Here, non-conservation of energy at collisions needs special attention.
It is natural that because of the non-deterministic nature of the dynamics, Lagrangian relations instead of Lagrangian submanifolds arise.
Still, these form a potential bridge for comparing our nondeterministic dynamics with the deterministic one of many-body motion with interactions.

\section{Notation} \label{sec:notation}
%
\subsection{Phase Flow}
%
We consider $n\geq1$\label{number of particles} 
particles of masses $m_i>0$\label{masses} at positions $q_i\in\bR^d$
in their configuration spaces $i\in N := \{1,\ldots,n\}$.
On the configuration space\label{total configuration space} $ \bR^{nd}$
of all particles we use the inner product
$\LA\cdot,\cdot\RA_\cM$
generated by the mass matrix
\[\cM := {\rm diag}(m_1,\ldots,m_n)\otimes \idty_d,\]
\beq
\LA\cdot,\cdot\RA_A:\bR^{nd}\times \bR^{nd}\to\bR \qmbox{,} 
\LA q,q'\RA_A :=  \LA q,A q'\RA,
\Leq{inner:product}
denoting the bilinear form for the 
matrix $A\in {\rm Mat}(nd,\bR)$, with
the canonical inner product $\LA \cdot,\cdot\RA$.
We set $\|q\|_\cM:=\sqrt{\LA q,q\RA_\cM}$.
As we consider dynamics that conserves the total momentum, without loss of generality 
we restrict ourselves to the {\em center of mass configuration space} 
\beq 
\textstyle M := \{q = (q_1,\ldots,q_n) \in \bR^{nd} \mid \sum_{k=1}^n m_k q_k=0\} 
\Leq{M:0}
of dimension $D := \dim (M) = (n-1)d$.
The {\em collision set} in configuration space is given by
\beq
\Delta := \{q\in M \mid q_i=q_j\mbox{ for some }i\neq j\in N\} \, .
\Leq{coll:set}
%
\subsection{Cluster Decomposition}
%
Non-collision singularity orbits experience an infinity of cluster changes, 
so that we have to set up
a notation that allows to describe them. So we consider the set partitions of 
$N=\{1,\ldots,n\}$, see, {\em e.g.\!} {\sc Aigner} \cite{Ai}, and {\sc Derezi\'{n}ski}
and {\sc G\'{e}rard}~\cite{DG}.
\begin{defi}\quad\label{def:meet:join}\\[-6mm]
\begin{enumerate}[$\bullet$]
\item
A {\bf set partition} (or {\bf cluster decomposition})
of $N$ is a set
$\cC:=\{C_1,\ldots,C_k\}$ of {\bf atoms} (or {\bf clusters})
$\es\neq C_\ell\subseteq N$ with
\[\textstyle
\bigcup_{\ell=1}^kC_\ell=N \qmbox{and}
C_\ell\cap C_m=\es \mbox{ for } \ell\ne m \, .\]
\item
The equivalence relation on $N$ induced by $\cC$ has equivalence 
classes denoted by $[\cdot]_\cC$ or simply $[\cdot]$.
\item
The {\bf partition lattice} $\cP(N)$ is the set of 
cluster decompositions $\cC$ of $N$, partially ordered by 
{\bf refinement}, that is
\[\cC=\{C_1,\ldots,C_k\} \preccurlyeq \{D_1,\ldots,D_\ell\} = \cD, 
\qmbox{if} C_m \subseteq D_{\pi(m)}\] 
for a suitable map
$\pi:\{1,\ldots,k\}\to\{1,\ldots,\ell\}$. 
Then $\cC$ is called {\bf finer} than $\cD$ and $\cD$ {\bf coarser} than $\cC$.
\item
The {\bf rank} of $\cC\in\cP(N)$ is the number $|\cC|$ of its atoms.
\item
We denote by
$\cC\wedge\cD$ the  {\bf meet} (the coarsest partition finer
than $\cC$ and $\cD$) and by $\cC\vee\cD$ the {\bf join} (the finest partition coarser
than $\cC$ and $\cD$) of $\cC, \cD\in\cP(N)$.
\item
The unique finest resp.\ coarsest elements of $\cP(N)$ are denoted by
\beq
\cC_{\min} := \big\{ \{1\},\ldots,\{n\} \big\}\qmbox{and}
\cC_{\max} := \big\{ \{1,\ldots,n\}\big\} \, ,
\Leq{fine:coarse}
and we set
\beq
\cP_\Delta(N) := \cP(N)\setminus\{\cC_{\min}\} \, . 
\Leq{cP:Delta}
\end{enumerate}
\end{defi}
A non--empty subset $C\subseteq N$ corresponds to the subspace\label{cluster subspace} 
\beq
\Delta^E_C:=\{q\in M\mid q_i = q_j\ \mbox{for}\ i,j\in C\} \, .
\Leq{def:MEC}
The superscript $E$ (which we often omit in the subsequent chapters)
refers to the fact that only coordinates
{\em external} to the cluster $C$ vary on this subspace.
For a partition $\cC$ we define the $\cC$--{\em collision} subspace
\[ \Delta_\cC \equiv\Delta^E_\cC := \l\{q\in M\mid q_i=q_j\ \mbox{if}\ [i]_\cC=[j]_\cC \ri\}\]
(this specialises to (\ref{def:MEC}) if we attribute to $C$ 
the partition $\cC$
which only consists of $C$ and one--element clusters).
We then have $\Delta^E_\cC=\bigcap_{C\in\cC}\Delta^E_{C}$.

The $\cM$--{\em orthogonal} (w.r.t.\ the inner product $\LA\cdot,\cdot\RA_\cM$ from 
(\ref{inner:product}))
projection $\Pi^E_C$\label{cluster projection} onto $\Delta^E_C$ is 
complemented by $\Pi_C^I:=\idty_M-\Pi^E_C$ with range \label{MCbot}
\beq \textstyle
\Di_C = {\rm ker}\l(\Pi^E_C\ri)
=\big\{q\in M\mid q_i=0\ \mbox{for}\ i\not\in C,\ \sum_{i\in C}\,m_iq_i =0\big\}.
\Leq{M:c:bot}
Here the superscript $I$ refers to the fact that only coordinates
{\em internal} to the cluster $C$ vary on this subspace.
Similarly for a partition $\cC$
the linear maps 
\beq
\Pi^E_\cC := \prod_{C\in\cC}\Pi^E_C
\qmbox{respectively} \Pi^I_\cC := \idty_M-\Pi^E_\cC=\sum_{C\in\cC}\Pi^I_C
\quad\mbox{on }M
\Leq{proj:E} 
are the $\cM$--orthogonal projections
onto $\Delta^E_\cC$ resp. $\Di_\cC$, with dimensions
\beqn
\dim(\Delta^E_\cC) &=& 
d\l(n-1-\sum_{C\in \cC}(|C|-1)\ri) =  d\,\big( |\cC|-1 \big)\ ,\nonumber\\
\dim(\Di_\cC) &=& \sum_{C\in \cC}\dim(\Di_{C}) 
= d\sum_{C\in \cC} ( |C|-1 ) 
= d \big(n-|\cC| \big) \, .
\label{dim:MC}
\eeqn
This implies $\Di_\cC = \bigoplus_{C\in \cC} \Di_{C}$,
since $\Delta^E_\cC=\bigcap_{C\in \cC}\Delta^E_{C}$. Moreover by (\ref{M:c:bot})
\beq
M =\Delta^E_\cC \oplus \Di_\cC =\Delta^E_\cC \oplus \bigoplus_{C\in \cC}\Di_{C}
\qquad \big(\cC\in \cP(N)\big)
\Leq{decomp}
are $\cM$--orthogonal decompositions. 

The following relations between collision subspaces derive from
meet and join of Def.\ \ref{def:meet:join}.
\beq
\Delta^E_{\cC\vee\cD} =\Delta^E_\cC\cap\Delta^E_\cD \qmbox{so that}
\Di_{\cC\vee\cD} = \Di_\cC + \Di_\cD \qquad
\big(\cC,\cD\in \cP(N)\big).
\Leq{part:perp}
Note that conversely we only have
\[
\Di_{\cC\wedge\cD} \subseteq \Di_\cC \cap \Di_\cD \qmbox{so that}
\Delta^E_{\cC\wedge\cD} \supseteq\Delta^E_\cC +\Delta^E_\cD \qquad
\big(\cC,\cD\in \cP(N)\big).
\]
\begin{example}[Collision Subspaces]
For $n:=4$, $\cC:=\{\{1,2\},\{3,4\}\}$ and $\cD:=\{\{1,3\},\{2,4\}\}$ 
we get $\cC\wedge\cD=C_{\min}$ so that
\[M=\Delta^E_{\cC\wedge\cD} \;\supsetneq \;\Delta^E_\cC + \Delta^E_\cD = 
\{q\in M\mid q_1+q_4=q_2+q_3\} \, . \hfill \tag*{$\Diamond$} \]
\end{example}
To have a less heavy notation we set
\[q^E_\cC:=\Pi^E_\cC(q)\qmbox{and}q^I_\cC:=\Pi^I_\cC(q)\qquad(q\in M),\]
and even omit the subscript $\cC$ when the context permits.
For a nonempty subset $C\subseteq N$ we define the \label{cluster mass} 
{\em cluster mass\,} and {\em cluster barycenter} of $C$ by
\[m_C := \sum_{j\in C} m_j \qmbox{and}
  q_C := \frac{1}{m_C} \sum_{j\in C} m_j q_j \, .\]
In particular $m_N$ equals the 
{\em total mass} of the particle system.
Then for the partitions $\cC\in \cP(N)$ the $i$--th component of the 
cluster projection is given by the barycenter
\beq
\l(q^E_\cC\ri)_i = q_{[i]_\cC} \qquad (i\in N)
\Leq{cl:bar}
of its atom. Similarly
\[ \l(q^I_\cC\ri)_i = q_i-q_{[i]_\cC} \qquad (i\in N).\]
is its distance from the barycenter. The {\em scalar moment of inertia}
\beq
J:M\to\bR \qmbox{,} J(q):= \eh \LA q,q\RA_\cM
\Leq{scalar:moment:of:inertia}
splits into the {\em cluster barycenter moment}
\beq
J^E_\cC := J\circ\Pi^E_\cC \qmbox{,} J^E_\cC(q)=\sum_{C\in\cC} 
\tfrac {m_{C}}{2} \LA q_C,q_C\RA
\Leq{cluster:barycenter:moment}
and the {\em relative moments of inertia} of the clusters $C\in\cC$
\[J_{C}^I := J\circ\Pi_{C}^I \mbox{ , } 
J_{C}^I(q) = \sum_{i\in C} \tfrac {m_i}{2} \|\l(q^I_\cC\ri)_i\|^2 = 
\frac{1}{2m_C} \sum_{i<j\in C} m_i m_j \LA q_i-q_j,q_i-q_j\RA ,\]
that is
\beq
J=J^E_\cC+J^I_\cC\qmbox{for} 
J^I_\cC(q) := \sum_{C\in\cC}J_{C}^I(q) = \eh \LA q^I_\cC,q^I_\cC\RA_\cM.
\Leq{J:decomp}
Denoting by $M^*$ the dual space of the vector space $M$, there are natural 
identifications $TM\cong M\times M,\ T^*M\cong M^*\times M$ of the tangent
space resp.\ phase space of $M$. This gives rise to the inner products
\[\LA \cdot,\cdot\RA_{TM}:TM\times TM\to\bR \qmbox{,} 
\LA (q,v),(q',v')\RA_{TM}:=\LA q,q'\RA_\cM+\LA v,v'\RA_\cM\]
and
\beq
\LA \cdot,\cdot\RA_{T^*M}:T^*M\times T^*M\to\bR \qmbox{,} 
\LA (p,q),(q',p')\RA_{T^*M}:= \LA q,q'\RA_\cM+\LA p,p'\RA_{\cM^{-1}}
\Leq{cotangent:scalar:product}
$\l(\mbox{with}\ \LA p,p'\RA_{\cM^{-1}}=\sum_{i=1}^n\frac{\LA p_i,p_i'\RA}{m_i}\mbox{ for 
the {\em momentum vector} } p=(p_1,\ldots,p_n)\ri)$.

The tangent space $TU$ of any subspace $U\subseteq M$ is naturally a 
subspace of $TM$. 
Using the inner product, we also consider $T^*U$ as a subspace of $T^*M$.

We thus obtain $T^*M$--orthogonal decompositions
\[T^*M =
T^*\Delta^E_\cC\oplus\bigoplus_{C\in\cC}T^*(\Di_C)\qquad
\big(\cC\in\cP(N)\big)\]
of phase space. 
With  
\[\widehat{\Pi}^I_\cC := \idty_{T^*M}-\widehat{\Pi}^E_\cC
=\sum_{C\in\cC}\widehat{\Pi}^I_C\]
the $T^*M$--orthogonal projections 
$\widehat{\Pi}^E_\cC, \, \widehat{\Pi}^I_\cC: T^*M \to T^*M$
\label{widehatPicC}
onto these subspaces are given by the {\em cluster coordinates}
\label{cluster coordinates}
\beq
(p^E,q^E) := \widehat{\Pi}^E_\cC(p,q) \qmbox{with} (p^E_i,q^E_i)
= \l(\frac{m_i}{m_{[i]}}p_{[i]}, q_{[i]}\ri)\quad(i\in N),
\Leq{cl:co}
and {\em relative coordinates}\label{relative coordinates}
\[(p^I,q^I):=\widehat{\Pi}^I_\cC(p,q) \qmbox{with} (p^I_i,
q^I_i) = (p_i-p^E_i , q_i-q^E_i) \quad (i\in N).\]
Here $p_C:=\sum_{i\in C}p_i$ is the {\em total momentum of the cluster}
$C\in \cC$. Unlike in (\ref{cl:bar}) we omitted the subindex $\cC$
in (\ref{cl:co}), but will include it when necessary.
\begin{lemma}\label{lem:sym}
The vector space automorphisms 
\beq
\l(\widehat{\Pi}^E_\cC,\widehat{\Pi}^I_\cC \ri):T^*M\longrightarrow
T^*\Delta^E_\cC\oplus\bigoplus_{C\in\cC}T^*(\Di_C)
\qquad\big(\cC\in \cP(N)\big)
\Leq{sy:trans}
are symplectic w.r.t.\ the natural symplectic forms on these cotangent bundles.
\end{lemma}
{\bf Proof.}
This follows from 
$T^*\big(\Delta^E_\cC\oplus\bigoplus_{C\in\cC}
\Di_C\big) = T^*\Delta^E_\cC\oplus\bigoplus_{C\in\cC}T^*(\Di_C)$.\hfill$\Box$\\[2mm]
{\em Total angular momentum}
\beq
L:T^*M\to\bR^d\wedge\bR^d \qmbox{,} L(p,q) = \sum_{i=1}^nq_i\wedge p_i
\Leq{total:ang:momentum}
and {\em total kinetic energy}
\[K:T^*M\to\bR \qmbox{,} K(p,q) \equiv K(p)=\eh\LA p,p\RA_{M^*} = \eh\LA p,p\RA_{\cM^{-1}} 
= \sum_{i=1}^n\frac{\LA p_i,p_i\RA}{2m_i}\]
(using a sloppy notation) both split into sums of {\em barycentric}
\beqno
L^E_\cC &:=& L\circ\widehat{\Pi}^E_\cC \qmbox{,} 
L^E_\cC(p,q) = \sum_{C\in\cC}q_C\wedge p_C \, ,\\
K^E_\cC &:=& K\circ\widehat{\Pi}^E_\cC \qmbox{,} K^E_\cC(p,q) = \sum_{C\in\cC}
\frac{\LA p_C,p_C\RA}{2m_C}
\eeqno
and {\em relative} terms for the clusters $C\in\cC$
\beqno
L_C^I &:=& L\circ\widehat{\Pi}_C^I \qmbox{,} 
L_C^I(p,q) = \sum_{i\in C}q^I_i\wedge p^I_i\\
K_C^I &:=& K\circ\widehat{\Pi}_C^I \qmbox{,} 
K_C^I(p,q) = \sum_{i\in C}\frac{\LA p^I_i,p^I_i\RA}{2m_i} \, .
\eeqno
That is,
\[L = L^E_\cC + \sum_{C\in\cC}L_C^I \qmbox{and} 
K = K^E_\cC + K^I_\cC \qmbox{with}K^I_\cC:= \sum_{C\in\cC}K_C^I \, .\]

For non-trivial partitions 
neither the cluster coordinates nor the relative coordinates are
coordinates in the strict sense.
Later, however, we need such coordinates
on the above-mentioned symplectic subspaces of phase space.

\begin{example}[Pair Clusters]\label{ex:1}
For each pair $i<j\in N$ we 
have the cluster decomposition $\cC$ containing the only nontrivial 
cluster $C:=\{i,j\}$ (these partitions will be important since for
$n>2$ they generate $\cP(N)$).
In this case
\[(q^I_\cC)_i = \frac{m_j}{m_i + m_j}(q_i - q_j)\qmbox{,}
(p^I_\cC)_i = \frac{m_jp_i - m_ip_j}{m_i + m_j} = - p^I_j \, ,\]
so that the $d$ components of
$q^I_C := (q^I_\cC)_i-(q^I_\cC)_j = q_i-q_j$ and of
$p^I_C := (p^I_\cC)_i$ \\
meet the canonical commutation relations,
\beq
J^I_\cC(q) = J^I_C(q) = \eh \frac{m_im_j}{m_i+m_j} \|q^I_C\|^2
\mbox{ and }
K^I_\cC(p) = K^I_C(p) =\eh \frac{m_i+m_j}{m_im_j} \|p^I_C\|^2 \, . 
\Leq{J:K:I}
Similarly  the components of $q_{[i]}$ and $p_{[i]}$
meet the canonical commutation relations, Poisson-commute 
with $q^I$ and $p^I$, 
and
$K^E_\cC(p,q) = \sum_{C\in\cC}\frac{\LA p_C,p_C\RA}{2m_C}$. 
\hfill$\Diamond$
\end{example}
Since for solutions of the Hamilton equations
the momentum $p_C:=\sum_{i\in C}p_i$ of the clusters $C\in\cC$ is 
related to the time derivative of the cluster barycenter (\ref{cl:bar})
by $p_C=m_C\dot{q}_C$, we get
\beq
J^E_\cC(\dot{q}) = K^E_\cC(p) \qmbox{and} J^I_\cC(\dot{q}) = K^I_\cC(p).
\Leq{K:J}

\begin{remark}[partition of configuration space]\quad\\
Because of (\ref{part:perp}) the subspaces $\Delta_\cC$ of $M$ 
generate a set partition of $M$ with the atoms
\beq
\Xi_\cC := \Delta_\cC\Big\backslash
\bigcup_{\cD\succneqq\cC}\Delta_\cD \qquad \big( \cC\in \cP(N) \big).
\Leq{Xi:Null}
As $\Xi_\cC$ is open in the vector space $\Delta_\cC$, by \eqref{dim:MC} 
it is a manifold with 
$\dim(\Xi_\cC) = d\,(|\cC|-1)$.
\hfill $\Diamond$
\end{remark}
\subsection{Radial coordinates} \label{sub:radial:coordinates}
For $q\in M\backslash\{0\}$ we can define the canonical coordinates 
$(R,Q) := (\|q\|_\cM,q/\|q\|_\cM)$,
\beq
P_R(p,q)= \frac{\LA p,q\RA}{\|q\|_\cM} \qmbox{and}
P_Q(p,q)= \|q\|_\cM \ p- \frac{\LA p,q\RA}{\|q\|_\cM}\cM q \, .
\Leq{P:RQ:map} 
We will show below in Theorem \ref{thm:finer}.5, that for non-deterministic trajectories 
$q:I\to M$ the numerator $\LA p,q\RA(t)$ of $P_R$ 
can be defined for all $t\in I$, unlike $p$ itself.
Then in absence of total collision for $t\in I\backslash \cT$
\begin{align}
\dot{Q} = \frac{\cM^{-1}P_Q}{R^2}&\mbox{ , }
\dot{P}_Q = -\frac{\|P_Q\|_{\cM^{-1}}^2}{R^2}\cM Q
\mbox{ , } \dot{R} = P_R\ \mbox{ and }\
\dot{P}_R = \frac{\|P_Q\|_{\cM^{-1}}^2}{R^3} \, .
\label{tr:ham:eq}
\end{align}
In particular $P_R$ is monotone increasing. Using $W:=P_Q/{R^2}$ instead of $P_Q$,
\[\dot{W} = - \|W\|_{\cM^{-1}}^2\cM Q -\frac{2P_R}{R}W\mbox{ , }
 \dot{Q} = \cM^{-1}W\mbox{ , }\dot{R} = P_R \mbox{ and } \dot{P}_R = R\|W\|_{\cM^{-1}}^2.\]
%
\section{Nondeterministic systems} \label{sec:nondeterministic}
%
In \cite{Bo} {\sc Bolotin} defined degenerate billiards as ones with scattering manifold
of codimension larger than one. Upon reflection, the change of momentum was assumed
to be perpendicular to its tangent space, the total energy being conserved.
In \cite{FKM}, scattering by a union of linear subspaces in configuration space 
was considered.
Here we further generalise these ideas by not demanding conservation of energy during 
collisions. 

\subsection{Definition and elementary properties}

The free flow on phase space $P=T^*M$ with canonical symplectic form $\omega$
is induced by the (purely kinetic) Hamiltonian 
\[K:P\to \bR\qmbox{,}K(p,q) := \eh\|p\|_{\cM^{-1}}^2\] 
and takes the usual form 
\beq
\Phi:\bR\times P\to P\qmbox{,}\Phi(t;p,q) = (p,q+\cM^{-1}p\,t) \,.
\Leq{free:flow} 

We now model the asymptotic motion of $n$ particles on $\bR^d$ 
by a nondeterministic motion where  between collisions 
the particles move with constant velocity,
that is via $\Phi$. 
These particles are allowed to have equal positions for non-trivial time intervals,
imitating close deterministic clusters. Spontaneous breakup of such clusters is allowed, too.
During collision, the total momentum (but not necessarily the total energy) of the colliding   
group is assumed to be constant.
\begin{defi}\label{def:ND}\quad\\
The {\bf nondeterministic system} of $n\in\bN$ particles with masses 
$m_1, \ldots, m_n > 0$\linebreak  
and in $d\in\bN$ dimensions is the set of {\bf (nondeterministic) trajectories}\linebreak 
$q\in C(I,M)$, having the following properties:
\begin{enumerate}[1)]
\item 
$I:= (T^-,T^+)$ for {\bf escape times} $-\infty \le T^- < 0 < T^+ \le \infty$.
If $T^\pm \neq \pm \infty$, the limits $\lim_{t\to T^\pm} q(t)$ do not exist in $M$.
\item 
With the {\bf collision set} $\mathrm{CS}:I\to \cP(N)$ defined by 
$q(t) \in \Xi_{\mathrm{CS}(t)}$, we set
\beq
\cT := \{t\in I\mid \mathrm{CS} \mbox{ is not locally constant at }t\} \, .
\Leq{T:nlc}
The momentum $p = \cM q'$ is constant in each interval of $I\backslash \cT$.
\item 
For all $\tau\in I$ the map $($see \eqref{Xi:Null}$)$
\[q_{\mathrm{CS}(\tau)}^E:I\to \Delta_{\mathrm{CS}(\tau)}^E\]
is continuously differentiable in an open interval $U\ni \tau$, and the external momentum
$U\ni t\mapsto p_{\mathrm{CS}(\tau)}^E(t) := \cM \, (q_{\mathrm{CS}(\tau)}^E)'(t)$ 
is locally constant at $\tau$.
\end{enumerate}
{\bf We use the shorthand $t\mapsto f_{\mathrm{CS}}(t)$ for
maps $t\mapsto f_{\mathrm{CS}(t)}(t)$.} 
\end{defi}
\begin{remarks}[nondeterministic system] \quad\\[-6mm] \label{rem:q:cS:0}
\begin{enumerate}[1.]
\item
Assumption 1) means that the nondeterministic trajectory is not the restriction of a 
nondeterministic trajectory on a strictly larger time interval.
\item  
Assumption 2) implies that  in each interval of $I\backslash \cT$ the particles 
in a cluster of $\mathrm{CS}(t)$ 
share the same position and move with the same constant velocity.
\item
By Assumption 3) the cluster centers move freely in the time intervals $U$.
\item
Note that the nondeterministic systems are invariant under time  translations
and reversible: for a trajectory $q\in C(I,M)$ and $t\in I=(T^-,T^+)$, 
\[q(\bullet-t):(T^--t,T^+-t)\to M\qmbox{and}q(-\bullet):(-T^+,-T^-)\to M\] 
are trajectories, too. Thus it suffices to prove properties for time zero and in the future 
direction. The orthogonal group $\mathrm{O}(d)$ acts diagonally on 
$M \subseteq (\bR^d)^n$ and
maps nondeterministic trajectories to  nondeterministic trajectories.
\item
For every trajectory $q\in C(I,M)$ we have the one-parameter family of 
trajectories:
\[q_\lambda\in C(I/\lambda,M)\qmbox{,}
q_\lambda(t)= q(\lambda t)\qquad(\lambda\in \bR^+).
\]
Additionally, rescaling by $q\mapsto \mu q$ is possible.
\hfill $\Diamond$
\end{enumerate}
\end{remarks}
\begin{lemma} \label{lem:T:closed}
$\mathrm{CS}$ is a Borel measurable map and $\cT$ a closed  Borel subset of $I$. 
\end{lemma}
\textbf{Proof:}
Measurability of $\mathrm{CS}$ follows, as the $\Xi_\cC\subseteq M$ 
are measurable and $q$ is continuous.
An accumulation point $\tau\in I\backslash \cT$ of $\cT$ cannot exist,
since then $\mathrm{CS}$ would not be locally constant at $\tau$ by definition \eqref{T:nlc}.
\hfill $\Box$\\[2mm]
Unlike ordinary billiards with energy conservation, nondeterministic systems are
projectable in the following sense:
\begin{lemma}[projections of nondeterministic trajectories] \quad\\[-6mm] 
\label{lem:projection}
\begin{enumerate}[1.]
\item 
Every orthogonal projection $\Pi_1:\bR^d\to \bR^d$ gives rise to an $\cM$--orthogonal 
projection
$\Pi_n:=(\Pi_1,\ldots, \Pi_1):\bR^{nd}\to \bR^{nd}$ with $\Pi_n(M)\subseteq M$. 
\item 
$\Pi_n$ maps trajectories to parts of  trajectories.
\end{enumerate}
\end{lemma}
\begin{remark}[extension of nondeterministic trajectories]\quad\label{rem:projection}\\
Note that we can always extend a trajectory on $(a,b)$, 
for which the positions and the momenta have a limit at the boundary points, 
by just letting the momenta be constant outside the interval and the positions 
evolve linearly.
\hfill $\Diamond$
\end{remark}
\textbf{Proof of Lemma \ref{lem:projection}:}
\begin{enumerate}[1.]
\item 
$\cM$--orthogonality of $\Pi_n$ 
follows from the definition \eqref{inner:product} of the inner product 
$\LA\cdot,\cdot\RA_\cM$ on $\bR^{nd}$. $\Pi_n$--invariance of $M$ is immediate
from \eqref{M:0}: For $q\in M$, 
$\sum_{k=1}^n m_k \Pi_1(q_k) = \Pi_1(\sum_{k=1}^nm_k q_k) = 0$.
\item 
First note that (unlike $\lim_{t \rightarrow T^{\pm}}q(t)$)  the limits 
$\lim_{t \rightarrow T^{\pm}} \Pi_n(q(t))$ may exist, a simple example being that of the trivial 
projection $\Pi_1 \equiv 0$. 
But we can prove that on the interval $I=(T^{-}, T^{+})$ the properties of 
Definition~\ref{def:ND} are satisfied. 
Therefor we define  $\widetilde{\mathrm{CS}}:I\to \cP(N)$ by 
$\Pi_n(q(t))\in \Xi_{\widetilde{\mathrm{CS}}(t)}$. Then we get a new collision set
\[ \widetilde{\cT}:= \{t\in I\mid \widetilde{\mathrm{CS}} \mbox{ is not locally constant at }t\} \, .
\]
We need to show that $\tilde{p} = \cM \Pi_n(q')$ is constant in each interval of $I \setminus \widetilde{\cT}$ and that for all $\tau\in I$ the map 
\[q_{\widetilde{\mathrm{CS}}(\tau)}^E:I\to \Delta_{\widetilde{\mathrm{CS}}(\tau)}^E\]
is continuously differentiable in an open interval containing $\tau$.
\\
The second statement is the simpler one. Note that for any time $t \in I$ we must have 
$\mathrm{CS}(t) \preccurlyeq \widetilde{\mathrm{CS}}(t)$. 
Denote with $\Pi_{\mathrm{CS}(t), \widetilde{\mathrm{CS}}(t)}: \Delta_{\mathrm{CS}(t)}^{E} 
\rightarrow \Delta_{\widetilde{\mathrm{CS}}(t)}^{E}$ 
the well-defined projection between these subspaces. 
Then for any $\tau \in I$ we have that $q_{\widetilde{\mathrm{CS}}(\tau)}^E = 
\Pi_{\mathrm{CS}(\tau), \widetilde{\mathrm{CS}}(\tau)} \circ q_{\mathrm{CS}(\tau)}^E$ is a 
composition of a linear projection and a differentiable map (in a neighborhood of $\tau$) and 
hence differentiable (in a neighborhood of $\tau$). \\
For  $\tilde{p} = \cM \Pi_n(q')$ being constant on any interval in $I \setminus \widetilde{\cT}$ it 
is sufficient that for any $t \in I \setminus \widetilde{\cT}$ there is a neighborhood of $t$ on 
which $\tilde{p}$ is constant. \\
Note that we neither can guarantee $\widetilde{\cT} \subseteq \cT$ nor 
$\cT \subseteq \widetilde{\cT}$. But if $q$ is differentiable and $p$ is constant in a 
neighborhood of some time $t \in I$, the same holds true for $\Pi_n(q)$ and $\tilde{p}$.
Thus we only have to consider times $t \in \left(I \setminus \widetilde{\cT}\right) \cap \cT$, i.e.\ 
where $\widetilde{\mathrm{CS}}$ is locally constant, but $\mathrm{CS}$ is not. Now as 
$\widetilde{\mathrm{CS}}$ is constant on some neighborhood $U_0 \ni t$ we know that 
$\Pi_n(q) = \Pi_n(q_{\widetilde{\mathrm{CS}}(t)}^{E}) 
= (\Pi_n(q))_{\widetilde{\mathrm{CS}}(t)}^{E}$ 
(note that it does not matter whether or not we project first and then take the cluster 
coordinates or vice versa). Therefore, 
$\Pi_n(q)$ is (on some possibly smaller neighborhood which we still denote with $U_0$)
continuously differentiable by the result from the first part of the proof. 
Thus we only need to show that the derivative is constant. 

Therefor, we need that $\cT$ is nowhere dense, compare Theorem~\ref{thm:finer} below. 
We know that (again possibly in a smaller neighborhood) $p$ is locally constant on every 
interval of $U_0 \setminus \cT$ and $\mathrm{CS}(s)$ is always at most finer for $s \in 
U_0$ than $\mathrm{CS}(t)$ which is finer than $\widetilde{\mathrm{CS}}(t)$. Hence, we 
can project $p(s)$ to $\tilde{p}(s)$ on intervals where $p(s)$ is defined in the same way 
as above to obtain $\tilde{p}(s)$. So $\tilde{p}$ is constant on these intervals, too, and as 
$\cT$ is nowhere dense and $\tilde{p}$ is continuous, it has to be constant on the whole 
interval $U_0$, proving the statement.
\hfill $\Box$
\end{enumerate}
\subsection{Moment of inertia}

%
Next we consider the {\bf scalar moment of inertia} {\em along a trajectory}
\[J\in C\big(I,[0,\infty)\big) \qmbox{,} J(t)=\eh \big\langle q(t),\cM q(t)\big\rangle \, ,\]
and similarly $J_\cC^E, J_\cC^I\in C\big(I,[0,\infty)\big)$ for $\cC\in \cP(N)$
(see \eqref{scalar:moment:of:inertia}),
\beq
J_\cC^I(t) = \eh \big\langle q^I_\cC(t),\cM q^I_\cC(t)\big\rangle \qmbox{,} 
J_\cC^E(t) = \eh \big\langle q^E_\cC(t),\cM q^E_\cC(t)\big\rangle \, .
\Leq{eq:J:decomposition}
We thus have $J=J_{\cC_{\min}}^E = J_{\cC_{\max}}^I$ and 
$J=J_{\cC}^E +J_{\cC}^I$ for all $\cC\in \cP(N)$.\\[2mm]
In order to show convexity of $J$, we use the following simple fact.
\begin{lemma} \label{lem:C0:convexity}\quad\\ 
A function $f\in C(U,\bR)$ on an open interval $U$ is convex if and only
if for every $x\in U$ there exists an open interval $U_x\subseteq U$ containing $x$ with
$f|_{U_x}$ convex.
\end{lemma}
\textbf{Proof:}\\[-0mm]
It is obvious that convexity on $U$ implies local convexity on the $U_x$.\\
In order to conversely show convexity of $f$, assuming local convexity, we consider a
compact interval $[y_0,y_1]\subseteq U$. As $x\in U_x$, 
it is covered by the $U_x$ for $x\in [y_0,y_1]$. 
By compactness there exists a finite subcover of $[y_0,y_1]$. 
By induction it suffices to consider
two of these open intervals, say $U_a$ and $U_b$ with non-empty overlap and prove
that the restriction of $f$ to $U_a\cup U_b$ is convex, too. It also suffices to consider
$x_0\in U_a\backslash U_b$,  and $x_1\in U_b\backslash U_a$ and to show 
(assuming $x_0<x_1$ without loss of generality) that
\beq
f(x_t)\le (1-t)f(x_0)+tf(x_1)\mbox{ with }
x_t := (1-t) x_0 + t x_1\mbox{ and } t \in (0,1) \, .
\Leq{x:t}
\begin{enumerate}[$\bullet$]
\item 
If $x_t\in U_a$, then there are $s_2>s_1>t$ with 
$x_{s_2}>x_{s_1}\in U_a\cap U_b$.
By assumption of convexity of $f|_{U_a}$, with the notation
${\rm sl}(t_1,t_2) := \frac{f(x_{t_2})-f(x_{t_1})}{x_{t_2}-x_{t_1}}$, the secants have slopes 
\[ {\rm sl}(0,t) \le {\rm sl}(t,s_1)\le {\rm sl}(s_1,s_2) \, .\]
Similarly, concerning $U_b$, we know that
\[ {\rm sl}(s_1,s_2) \le {\rm sl}(s_1,1)\le {\rm sl}(s_2,1).\]
Combining these, we infer from ${\rm sl}(t,s_1)\le {\rm sl}(s_1,1)$ that
${\rm sl}(t,s_1)\le {\rm sl}(t,1)$
and obtain the convexity condition for $x_t$:
\[\textstyle
{\rm sl}(0,t)\le {\rm sl}(t,s_1)\le {\rm sl}(t,1)\qquad(0<t<1). \]
This is equivalent to \eqref{x:t}.
\item 
The case $x_t\in U_b$ is similar.
\hfill $\Box$
\end{enumerate}
To show that  $\cT$ is countable, we decompose $\cT$ into the disjoint union
\beq \textstyle
\cT=  \bigcup_{\cC\in \cP(N)} \cT_\cC\ \mbox{, with }\
\cT_\cC := \{t\in \cT\mid \mathrm{CS}(t)\in\cC\} \, .
\Leq{cT:decomposition}
\begin{theorem}[moment of inertia]\quad  \label{thm:finer} \\[-6mm]
\begin{enumerate}[1.]
\item 
For every $\tau\in I$ there is a maximal open interval $U_\tau\subseteq I$ containing $\tau$
with 
\[\mathrm{CS}(t) \preccurlyeq \mathrm{CS}(\tau) \qquad (t\in U_\tau).\]
\item 
$J$ is convex. More precisely, for every $\tau\in I$ and 
$\mathrm{CS}(\tau) = \{C_1,\ldots, C_k\}$ the functions
$J_{C_\ell}^E$ and $J_{C_\ell}^I$ $(\ell=1,\ldots,k)$ are convex on $U_\tau$.
\item 
Accumulation points of $\cT_\cC$ in $I$ are contained in the $\cT_\cD$ with
$\cC\precneqq\cD$.\\ 
So $\cT$ is countable and nowhere dense.
\item 
The velocity $q':I\setminus \cT \to M$ is locally bounded in $I$.\\
So for compact intervals $C\subseteq I$ the length of the curve $q|_C$ is finite.
\item 
$J$ is continuously differentiable, with 
\beq
J'(t) = \big\langle q(t) , p_{\mathrm{CS}}^E(t) \big\rangle \qquad\big( t\in I\big).
\Leq{derivative:of:J}
\item 
The distributional derivative of $J' = \langle q_{\mathrm{CS}}^E , p_{\mathrm{CS}}^E\rangle$ 
is a Radon measure with density 
$\langle p_{\mathrm{CS}}^E , p_{\mathrm{CS}}^E\rangle_{\cM^{-1}}$. 
\end{enumerate}
\end{theorem}
\textbf{Proof:} 
\begin{enumerate}[1.]
\item 
The existence of an open interval $\widetilde{U}_\tau\ni \tau$ with 
$\mathrm{CS}(t) \preccurlyeq \mathrm{CS}(\tau)$ for all  $t\in \widetilde{U}_\tau$
follows as the $\Xi_\cC\subseteq \Delta_\cC^E$ 
are relatively open and $q$ is continuous.
Then the union $U_\tau$ of these $\widetilde{U}_\tau$ is open.
\item 
To show that $J$ is convex, by Lemma \ref{lem:C0:convexity} it suffices to prove that
$J$ is convex on $U_\tau $, for all $\tau\in I$. Using  for 
$\cC=\{C_1,\ldots, C_k\}:= \mathrm{CS}(\tau)$
the decomposition $J = J_{\cC}^E +J_{\cC}^I$, see \eqref{eq:J:decomposition}, 
we note that 
$J_{\cC}^E|_{U_\tau}$ is convex, since it is the restriction of a quadratic polynomial with
leading coefficient $\eh \|\dot{q}_\cC^E(\tau)\|_\cM^2\ge 0$.

We show that $J_{\cC}^I = \sum_{\ell=1}^k J_{C_\ell}^I$ is convex, by considering the
closed subsets $V_\ell:= \{t\in U_\tau\mid J_{C_\ell}^I(t)=0\}$. If always
$V_\ell=U_\tau$, we are done. Otherwise we choose an $\ell$ violating the condition
and restrict to the corresponding subsystem $N' := C_\ell$ of $|N'|\le n$ particles.
Since  $U_\tau\backslash V_\ell$ is the disjoint union of non-empty open intervals,
if the restriction of $J_{C_\ell}^I$ to $U_\tau$ were not convex, there would exist such an 
interval  $I' \subseteq U_\tau\backslash V_\ell$ on which $J_{N'}^I$ is not convex.
But for all $\tau'\in I'$, $\mathrm{CS}_{N'}(\tau')$ is strictly finer than $N'$.

Iterating this argument at most $n$ times, we end up with cluster of one particle,
for which the internal moment of inertia vanishes. This proves that
$J_{\mathrm{CS}(\tau)}^I$ is convex on $U_\tau$. 

So $J$, too, is convex on $U_\tau$.
Lemma \ref{lem:C0:convexity}  then shows that $J$ is convex on its domain of definition $I$. 
\item
To show that  $\cT$ is countable, we use the decomposition \eqref{cT:decomposition}
of $\cT$ into the disjoint union of $\cT_\cC$.
As $q(t)\in \Delta_\cC^E$ for $t\in \cT_\cC$ and $\Delta_\cC^E\subseteq M$
is closed, we have $q(\tau)\in \Delta_\cC^E$ for an accumulation point $\tau$ of $\cT_\cC$.

So $\cC \preccurlyeq \cD$. But $\cC \neq \cD$, since if $q(\tau)\in \Xi_\cC$, it would have 
a positive distance to
all $\Delta_\cE^E$ with $\cE\neq \cC$ not finer than $\cC$. By Part 2)
the trajectory has to move
to such $\Delta_\cE$ before coming back to $\Xi_\cC$. This would make the trajectory 
discontinuous at time $\tau$.

Each $\cT_\cC$ is nowhere dense, since to return to $\Xi_\cC$ needs a positive time.
As $\cP(N)$ is finite, $\cT$ is nowhere dense, too.
\item
For $\cC:= \mathrm{CS}(\tau)$, 
\begin{enumerate}[$\bullet$]
\item 
the external velocity $(q^E_\cC)'$ is constant on $U_\tau$, 
using Item 1.\ of the theorem. 
\item 
So we consider the internal velocity $(q^I_\cC)'$.
Its radial component is monotone increasing, since by the Cauchy-Schwarz inequality
\beq
\textstyle
\frac{d}{dt}\frac{\langle q^I_\cC(t)\, ,\, p^I_\cC(t)\rangle}{\|q^I_\cC(t)\|_\cM}
= \frac{\|p^I_\cC(t)\|_{\cM^{-1}}^2\|q^I_\cC(t)\|_\cM^2\,
-\, \langle q^I_\cC(t)\,,\,p^I_\cC(t)\rangle^2} {\|q^I_\cC(t)\|_\cM^3} \ge 0
\quad(t\in U_\tau\backslash \cT),
\Leq{rad:comp}
and since $t\mapsto \langle q^I_\cC(t) , p^I_\cC(t)\rangle$ is continuous, too, 
at the countable set $U_\tau\cap \cT$.
To show the last assertion, we do an induction on the rank 
$n-|\cC|$ of $\cC\in \cP(N)$.

For the $\cC = \mathrm{CS}(\tau)$ of minimal rank within $\mathrm{CS}(I)$, 
Part 3) of the theorem implies that $\tau$ is not an accumulation point of $\cT$.
In fact, $U_\tau\cap \cT$ contains at most one point $\tau'$ besides $\tau$.
This occurs if on the compact interval with endpoints $\tau$ and $\tau'$, we have
$ \mathrm{CS}(t)=\cC$.

So $(q^I_\cC)'$ is bounded on $U_\tau$. 
In the induction step we consider $q'$ near points $\tau\in\cT$ that are accumulation points of 
$\cT$ and use that on any compact interval in $U_\tau$ that does not contain $\tau$,
$(q^I_\cC)'$ is bounded.

By reversibility we can consider the motion on the interval
\[U_\tau^- := \{t\in U_\tau\mid t \le \tau\} \, .\]
As $q^I_\cC(\tau) = 0$, by the above
$\langle q^I_\cC(t) , p^I_\cC(t)\rangle$ is negative on $U_\tau^-$.
So the radial component \eqref{rad:comp} 
is monotone decreasing in absolute value and thus bounded on $U_\tau^-$.
\item 
If $U_\tau^-\cap \cT = \{\tau\}$, then $p^I_\cC$ is constant on  $U_\tau^-\backslash\{\tau\}$
and we are done.
Otherwise  $(U_\tau^-\cap \cT)\backslash\{\tau\}\not=\emptyset$, and we 
consider times $t\in U_\tau^-\backslash\{\tau\}$ larger than some such element.
For such $t$ there exist nearest $t^\pm\in U_\tau^-\cap \cT$
with $t^- < t < t^+\le \tau$, since $U_\tau^-\cap \cT$ is closed.
\item 
We then know that 
$\mathrm{CS}(t^-)\succneqq\mathrm{CS}(t)\precneqq\mathrm{CS}(t^+)$ and
that $\mathrm{CS}(t^-)$ and $\mathrm{CS}(t^+)$ are not comparable.\\

If $t^+<\tau$, then for the relative position 
\[ r(t) := q_{\cC}^I(t) = q(t) - \big[\,q(\tau) + (t-\tau) (q_\cC^E)' (\tau)\,\big] \, ,\] 
the vectors $r(t^-)$ and $r(t^+)$ span a two-dimensional subspace $S$ of $\Delta_\cC^I$,
and $S$ contains $r(\tau)=0$,
\[\textstyle
r(t) = \frac{t^+ - t}{t^+-t^-}r(t^-) + \frac{t - t^-}{t^+-t^-}r(t^+) \qmbox{and} 
r'(t) = q'(t) - (q_\cC^E)'(\tau) = \frac{r(t^+)-r(t^-)}{t^+-t^-},\] 
see Figure \ref{fig:lemma:bounded}.
As the sum of angles of a triangle equals $\pi$, for the angle 
$\alpha:= \measuredangle \big(\mathrm{span}(r(t^-)),\mathrm{span}(r(t^+))\big)$ at $r(\tau)$
the angles $\beta\in(0,\pi/2)$ and $\gamma$ between
$r'(t)$ and the perpendiculars to $\mathrm{span}(r(t^-))$ and $\mathrm{span}(r(t^+))$, 
respectively are related by
\[\gamma = \beta-\alpha.\]
Thus if we had $\beta\le\alpha$ (like in Figure \ref{fig:lemma:bounded}), after time $t^+$ the 
motion would cease to be inwards, contrary to our assumption.

This observation leads to upper bound for the norms of $r'(t)$ and for 
$(r_{\mathrm{CS}}^E)'(t^+)$ in terms of the norm of $(r_{\mathrm{CS}}^E)'(t^-)$.
Quantitatively, 

\beq
\|r'(t)\|_\cM = \frac{\|(r_{\mathrm{CS}}^E)'(t^-)\|_\cM}{\sin(\beta)}
\stackrel{!}{<} \frac{\|(r_{\mathrm{CS}}^E)'(t^-)\|_\cM}{\sin(\alpha)}
\Leq{rp:norm}
and
\beqno
\|(r_{\mathrm{CS}}^E)'(t^+)\|_\cM &=& \|r'(t)\|_\cM\sin(\gamma) 
= \|(r_{\mathrm{CS}}^E)'(t^-)\|_\cM\textstyle \frac{\sin(\beta-\alpha)}{\sin(\beta)}\\
&< & \|(r_{\mathrm{CS}}^E)'(t^-)\|_\cM \, .
\eeqno
\begin{figure}[h]
\centerline{\includegraphics[width=60mm]{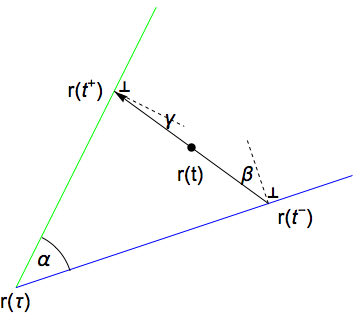}}
\caption{The plane $S$. $\gamma = \beta-\alpha$}
\label{fig:lemma:bounded}
\end{figure}
\end{enumerate}
The angle $\alpha$ in \eqref{rp:norm} is bounded below by the minimum of
the minimal positive (Jordan) angle 
between subspaces $\Delta_{\cC_1}, \Delta_{\cC_2}$
for non-comparable $\cC_1,\cC_2\in \cP(N)$.\\[2mm]
\noindent
We now prove the second statement of Part 4), finiteness of the length of the curve $q|_C$
for $C=[a,b]\subseteq I $. The length $\mathrm{Length}(q,C)\in [0,\infty]$
of $q$ on $C$ is given by
\[\mathrm{Length}(q,C):=
\sup\Big\{\sum_{k=1}^m\|q(t_k)-q(t_{k-1})\|_\cM\ \Big|\ a=t_0<\ldots < t_m=b\Big\} \, .\] 
Consider the family of $U_\tau$ for those $\tau\in C$ with $\mathrm{CS}(\tau)$ having 
minimal rank. By Theorem \ref{thm:finer}.1, 
this family is an open cover of the compact $C$. So it has a finite subcover.

So by the above, the continuous map $q':I\setminus \cT \to M$ is bounded on 
$C\backslash \cT$. Since
$\cT$ is countable, $q'$ is Lebesgue integrable on $C$ and 
$\mathrm{Length}(q,C) =  \int_C \|q'(t)\|_\cM\, \mathrm{d}t < \infty$. 
Thus $q'$ is locally rectifiable.
\item
By Remark \ref{rem:q:cS:0}, $q = q_{\mathrm{CS}}^E$ so that $J=J_{\mathrm{CS}}^E$ with
\[J_{\mathrm{CS}}^E(t) = 
\eh  \big\langle q_{\mathrm{CS}}^E(t),\cM q_{\mathrm{CS}}^E(t)\big\rangle
=  \eh  \big\langle q(t),\cM q_{\mathrm{CS}}^E(t)\big\rangle \, .\]
However, as we defined $q_{\mathrm{CS}}^E(t)$ as a shorthand for 
$q_{\mathrm{CS}(t)}^E(t)$, differentiability of $J$ at $\tau\in \cT$
does not directly follow from that formula.
Instead we use the decomposition $J(t) = J_{\cC}^E(t)+ J_{\cC}^I(t)$ for $t\in U_\tau$, with
$\cC:= \mathrm{CS}(\tau)$. 

The external part $J_{\cC}^E$ is the restriction of a quadratic 
polynomial and is thus continuously differential. 

We have $J_{\cC}^I(\tau)=0$, and by Item 4)
for any compact interval $C\subseteq U_\tau$ containing $\tau$, $\|(q_\cC^I)'\|_\cM$ is 
bounded, say by $k>0$. So 
\[\big| (J_{\cC}^I)'(\tau) \big| = \lim_{t\to 0} \frac{J_{\cC}^I(\tau+t)}{|t|}
=\eh \lim_{t\to 0} \frac{\|\int_\tau^{\tau+t} (q_\cC^I)'(s) \,\mathrm{d}s\|_\cM^2}{|t|}
\le \eh k^2\lim_{t\to 0} \frac{t^2}{|t|} = 0 \, . \] 
So the (ordinary) derivative of $J$ exists and has the form \eqref{derivative:of:J}.
\item
The Radon measure is absolutely continuous with respect to Lebesgue measure, since
\begin{enumerate}[$\bullet$]
\item 
its pure point part vanishes, as $J'$ is continuous by Part 2),
\item 
its singular continuous part vanishes, as by Part 4) $\cT$ is countable.
\end{enumerate}
As the second term in the distributional derivative
\[\langle (q_{\mathrm{CS}}^E)' , p_{\mathrm{CS}}^E\rangle
+\langle q_{\mathrm{CS}}^E , (p_{\mathrm{CS}}^E)'\rangle\]
is supported on $\cT$ and thus vanishes, its density is given by the first term.
\hfill $\Box$
\end{enumerate}
\begin{remark}[convexity of $J$ and properties of $\cT$]\quad\label{rem:J:T}\\ 
$J$ being a convex function, its 
second derivative exists almost everywhere, due to a general result
going back to Alexandrov. As $J''$ exists and is locally constant on $I\backslash \cT$,
this already implies that $\cT$ is of Lebesgue measure zero.\\
However, convexity of $J$ alone does not neither imply that $\cT$ is countable nor that
it is nowhere dense.
For a convex function $f:\bR\to \bR$ define $\cT$ as the set of points where $f''$ does not
exist.
\begin{enumerate}[$\bullet$]
\item 
A counterexample to countability of $\cT$ is the integral $f$ 
of the continuous, monotone increasing Cantor function. 
\item 
The second integral of the distribution
$\sum_{p/q\in \bQ}q^{-a}\delta_{p/q}$,
with $\mathrm{gcd}(p,q) = 1$ and $a > 2$ is an example of an $f$ 
for which $\cT$ is dense.
\hfill $\Diamond$
\end{enumerate}
\end{remark}
When $q(T_{\mathrm{coll}}) = 0$ for $T_{\mathrm{coll}}\in \cT$, 
we call $T_{\mathrm{coll}}$ a \textbf{total collision time}.\\ 
So $J(T_{\mathrm{coll}})=0$, 
and by convexity of $J$ (Theorem \ref{thm:finer}) there can be at most two 
such $T_{\mathrm{coll}}$ (the larger one corresponding to ejection).
\begin{remark}[accumulation points of $\cT$]\quad\label{rem:accumulation}\\
By Part 3 of Theorem \ref{thm:finer}, $\cT$ is countable and nowhere dense.
However, this does not mean that $\cT\subseteq I$ could only accumulate at
the escape times $T^\pm$.\\
A simple counterexample is the case of $n=3$ particles with equal masses 
in $d=1$ dimension.
Here the center of mass configuration space $M$ is two-dimensional. 
If we assume that the inner particle is keeping to be the same,
the motion corresponds to one in the cone $C=\{q\in M\mid q_1\le q_2\le q_3\}$ 
of opening angle $\pi/3$, with 
the total collision point $0$ as its tip. We assume that $J'(0)<0$. 
Then the trajectory $q:I\to C$ can be reflected successively by one and 
then the other ray forming $\partial C$. As long as these reflections at times $t_k$ are ingoing
($J'(t_k) < 0$), then one is free to choose the next collision with the other ray so that 
$J'(t_{k+1}) < 0$, too. By iterating one obtains a strictly increasing sequence of times 
$(t_k)_{k\in \bN}$.
The simplest choice is self-similarity of the trajectory. 
Then the outgoing angles $\alpha_k$ at these rays are all equal to some $\alpha$, which 
can be chosen in the interval $(\pi/3,\pi/2)$.

This leads to a decrease in distance from total collision and in speed which both are 
exponential in $k$:
By elementary geometry and the constancy of tangential velocity during collision 
we conclude that
\[\|q(t_{k+1})\|_\cM = c\,\|q(t_k)\|_\cM \qmbox{and}
\|q'(t_{k+1}^+)\|_\cM = s\,\|q'(t_k^+)\|_\cM\qquad(k\in \bN)\] 
for  $c := \cos(\alpha)/\cos(\alpha-\pi/3) \in(0,\eh)$ and 
$s := \sin(\alpha-\pi/3)/\sin(\alpha) \in(0,\eh)$.\\ 
If $\alpha = 5\pi/12$, then $c = s$ and thus $t_{k+1} - t_k$ is constant. More generally,
\begin{enumerate}[$\bullet$]
\item 
for $\alpha\in (\pi/3, 5\pi/12]$ the trajectory reaches the 
total collision point $0$ only in the limit $\lim_{k\to\infty} t_k = +\infty$.
\item 
Instead, for $\alpha\in (5\pi/12, \pi/2)$ the time $\lim_{k\to\infty} t_k$ of total collision is finite. 
\end{enumerate}
Thus we cannot assume that $\cT$ has a monotone enumeration 
$\{t_n\mid t_n < t_{n+1}\}$.~$\Diamond$
\end{remark}
Next we examine the asymptotics of the motion near the escape time $T^+$
(the one for escape time  $T^-$ following by reversibility). 
As the lattice $(\cP(N),\preccurlyeq)$ is complete,
the set partitions
\beq
\cC_{\sup} := \limsup_{t\nearrow T^+} \mathrm{CS}(t)\qmbox{and} 
\cC_{\inf} := \liminf_{t\nearrow T^+} \mathrm{CS}(t)
\Leq{C:sup:inf}
are defined, with $\cC_{\inf} \preccurlyeq \cC_{\sup}$.
The atoms of $\cC_{\sup}$ correspond to groups of particles that (as $t \nearrow T^+$) 
interact over and over again with one another, directly by collision or indirectly, 
via chains of colliding particles.\\
The atoms of $\cC_{\inf}$ correspond to groups of particles that stay together
as $t\nearrow T^+$.
\begin{lemma}  \label{lem:accu}
$\cC_{\inf} = \cC_{\sup}$ if and only if escape time 
$T^+$ is not an accumulation point of $\cT$ (which then implies $T^+ = +\infty$). 
\end{lemma}
\textbf{Proof:}
This is obvious from the definition \ref{def:ND} of $\cT$.
\hfill $\Box$\\[2mm]
As the partition lattice $\cP(N)$ is finite, there is a minimal time $t_{\min} \in [0,T^+)$
with 
\[  \sup\{ \mathrm{CS}(t)\mid t\in [t_{\min},T^+)\} = \cC_{\sup} 
\qmbox{and} 
\inf\{ \mathrm{CS}(t)\mid t\in [t_{\min},T^+)\} = \cC_{\inf} \, .\]
This will allow us to decompose the nondeterministic dynamics near $T^+$ by 
separately considering the subsystems corresponding to the clusters of $\cC_{\sup}$.

We call a nondeterministic trajectory $q$ {\em expanding} at some time $t$ if $J'(t)>0$.
However, this is not the only possible formalisation of that word.
The function 
\[ \mathrm{rad}:I\to [0,\infty)\qmbox{,} \mathrm{rad}(t) := \max\{\|q_k(t)\| \mid k\in N\} \]
is obviously continuous. 
It is even differentiable in every subinterval of $I \backslash \cT$ except maybe for finitely 
many points where some other particle attains the maximum.
Moreover, like $J$ it is convex: $\mathrm{rad}'(t_1)\ge \mathrm{rad}'(t_0)$ for all 
$t_1\ge t_0\in I\backslash \cT$,
since during collision involving particles $k$ realising that radius the total momentum 
is preserved. \\
We call $q$ {\em radius-expanding} at some time $t$ if $\mathrm{rad}'(t)>0$.
The following lemma relates the two forms of expansion.
\begin{lemma}[expansion and radius-expansion]\quad\\[-6mm]  
\begin{enumerate}[1.]
\item 
$J/m_N\le  \eh \mathrm{rad}^2 \le J/m_{\min}$.
\item 
If a nondeterministic trajectory $q$ is expanding at time $0$, it is radius-expanding
for all times larger than $\sqrt{m_N/m_{\min}} \, \frac{2J(0)}{J'(0)}$. 
\item 
Conversely, if a nondeterministic trajectory $q$ is radius-expanding at time $0$, it is 
expanding for all times larger than 
$\sqrt{m_N/m_{\min}} \, \frac{\mathrm{rad}(0)}{\mathrm{rad}'(0)}$.

\end{enumerate}
\end{lemma}
\textbf{Proof:}
\begin{enumerate}[1.]
\item 
This follows from the norm inequalities
$\|\cdot \|_\infty\le \|\cdot \|_2\le \sqrt{n} \, \|\cdot \|_\infty$ in $\bR^n$.
\item
With the notation from Subsection \ref{sub:radial:coordinates},
 \[ \textstyle
J(t) =\eh R(t)^2\ge \eh (R(0)+P_R(0)t)^2 = J(0) \big(1+\frac{J'(0)}{2J(0)} t\big)^2.\] 
 So, using Part 1., for $t\ge \sqrt{m_N/m_{\min}} \frac{2J(0)}{J'(0)}$ we have 
$\mathrm{rad}(t) \ge \mathrm{rad}(0)$, since
\[\eh \mathrm{rad}^2(t) \ge J(t)/m_N\ge J(0)/m_{\min}\ge  \eh \mathrm{rad}^2(0) \, .
 \]
\item 
$J(t)\ge \eh m_{\min}\, \mathrm{rad}(t)^2\ge 
\eh m_{\min} \, \mathrm{rad}(0)^2   \big(1+\frac{\mathrm{rad}'(0)}{\mathrm{rad}(0)}t \big)^2$. 
So, using Part 1., for $t\ge \sqrt{m_N/m_{\min}} \,\frac{\mathrm{rad}(0)}{\mathrm{rad}'(0)}$ 
we have $J(t)\ge J(0)$.
\hfill $\Box$
\end{enumerate}
Note that both $\lim_{t\nearrow T^+}\mathrm{rad}(t) = \lim_{t\nearrow T^+}J(t)=\infty$ for finite 
escape time $T^+$.

\subsection{Exponentially expanding systems}

Next we discuss the trajectories $q\in C(I,M)$ with  $I= (T^-,T^+)$ for which 
the escape time $T^+$ equals $\sup ( \cT)$. This is only possible if $|\cT|$ is (countably) 
infinite, which can only occur if $n\ge3$. 
If $T^+< \infty$, we must have $\limsup_{t\nearrow T^+}J(t)=\infty$.
As $J$ is convex (Theorem \ref{thm:finer}), this implies $\lim_{t\nearrow T^+}J(t)=\infty$,
which is a weak form of expansion.
\begin{example}[The case of three particles]\quad \label{ex:3:particles}\\
This case was analysed in some detail in \cite{Kn}.
It is easier than the one for $n\ge4$ since the three subspaces  
$\Delta_{\{\{1,2\},\{3\}\}}$,
$\Delta_{\{\{1,3\},\{2\}\}}$ and $\Delta_{\{\{2,3\},\{1\}\}}$ intersect pairwise only at 
$\Delta_{\{\{1,2,3\}\}} =\{0\}\in M$. Thus they are separated for $J>0$. Physically
this leads to a {\em messenger particle} moving back and forth between the other two
particles. So $\cT$ has no accumulation points.
It was shown in \cite{Kn} that the angular momenta of all three particles are 
uniformly bounded in their center of mass system 
and that the motion is asymptotically on a straight line.
However, the individual angular momenta become unbounded 
if the center of mass is not the origin. 
Moreover, and this is the most important difference to $n \ge 4$, assuming $J'(t_0)>0$,
for  a $c>1$ one has 
\[J(t_k)\ge c^k J(t_0)\] 
for collision times $t_{k}>t_{k-1}\in \cT$.

When comparing the non-deterministic dynamics with, say, celestial mechanics, one should keep in mind 
that the non-conservation of the energy of the former corresponds to the non-conservation
of the exterior part of the energy of clustering particles of the latter. So $n=3$ was used
to model the deterministic motion of {\em four} bodies with a tight binary.
\hfill $\Diamond$
\end{example}
Unlike in this example, for $n\ge4$ non-comparable pairs of subspaces  of $M$ may
have a non-trivial intersection. An example is  the intersection of
$\Delta_{\{\{1,2\},\{3\},\{4\}\}}$ and $\Delta_{\{\{1\},\{2\},\{3,4\}\}}$ which equals
$\Delta_{\{\{1,2\},\{3,4\}\}}$. 
To gain control of the motion, we observe that for $n=3$ and $T^+=\sup(\cT)$ 
by Lemma \ref{lem:accu} for all $k\in\bN$ one has 
$\mathrm{CS}(t_k)\vee \mathrm{CS}(t_{k+1})=\cC_{\sup}$.
It is this property that we generalise now to $n\ge4$.\\[1mm]

In the non-trivial case $\cC_{\inf}\neq \cC_{\sup}$ in \eqref{C:sup:inf} 
we now analyse in more detail
the collision set $\mathrm{CS}:I\to \cP(N)$ on $[t_{\min},T^+)$.
By performing a time translation by $t_{\min}$, we assume that we have
\[  \cC_{\inf} \preccurlyeq \mathrm{CS}(t) \preccurlyeq \cC_{\sup}\qquad \big(t\in[0,T^+)\big). \]
Beginning with $t_0:=0$ and with $\cD_0:=\mathrm{CS}(t_0)\precneqq \cC_{\sup}$, 
we inductively define minimal times  
\[t_k\in  (t_{k-1},T^+)\mbox{ with }\cD_k 
:= \sup \mathrm{CS}\big([t_0,t_{k}]\big) \,\succneqq\, \cD_{k-1} \, . \]
So $\cD_k = \sup \{ \mathrm{CS}(t_0),\ldots,\mathrm{CS}(t_k)\}$.
Denote by $k_{\max}\in N$ the maximal such $k$. Then 
\beq
\cD_{k_{\max}}= \cC_{\sup} \qmbox{and}k_{\max}\le |\cC_{\inf}|-|\cC_{\sup}|\le n-1 \,.
\Leq{kmax}
If $\cC_{\sup} \precneqq \{N\}$, then for times in $[0,T^+)$ the particles
corresponding to different atoms of $\cC_{\sup}$ do not collide, and we can consider
these subsystems separately. Of these there must exist at least one with escape time 
$T^+$. Thus we assume for the moment that  $\cC_{\sup} = \{N\}$.
Additionally we exclude total collision, that is $q(t)=0$ for some $t\in I$.
As $J$ is convex (Theorem \ref{thm:finer}.2), this is only technical.

Using the notation from Subsection \ref{sub:radial:coordinates},
for the radial velocity (which is a continuous function of time $t$) we thus get
\begin{align}
P_R(t) &= P_R(0)+\int_0^t R(s)\|W(s)\|_{\cM^{-1}}^2\,\mathrm{d}s\nonumber\\ 
&= P_R(0)+\int_0^t 
\big(\textstyle R(0) + \int_0^sP_R(\tau)\,d\tau\big)\,\|W(s)\|_{\cM^{-1}}^2\,\mathrm{d}s \, .
\label{PR:PR0}
\end{align}
Before we can formulate the main theorem, we need a simple lemma:
\begin{lemma}[long distance on the sphere $\bS$]\label{lem:long} \quad\\ 
There is a $k\in\{1,\ldots,k_{\max}\}$ such that
for the interval $C:=[t_{k-1}, t_{k}]$,  the length of the trajectory 
on the unit sphere $\bS\subseteq M$ is bounded below by
\beq
\mathrm{Length} \big(Q,[0, t_{k_{\max}}]\big)
\,\ge\, \textstyle \sum_{\ell=2}^{k_{\max}}{\rm dist}_\bS\big(Q(t_{\ell-1}),Q(t_\ell)\big) \,\ge\, D 
\Leq{dist:lob} 
and
\beq
\mathrm{Length}(Q,C)\,\ge\, {\rm dist}_\bS\big(Q(t_{k-1}),Q(t_k) \big) \,\ge\, \frac{D}{n}
\Leq{dist:lob:k} 
for
\beq
D := \min_{q\in \bS} \max \, \{ {\rm dist}_\bS(q,\Delta_{\cD_\ell} \cap \bS) \mid 
\ell = 0, \ldots, k_{\max} \}
> 0 \, .
\Leq{def:D} 
Here ${\rm dist}_\bS$ denotes the minimal angular distance on the unit sphere $\bS$. 
\end{lemma}
\textbf{Proof:}
\begin{enumerate}[$\bullet$]
\item 
As the $\Delta_{\cD_\ell}\cap \bS$ are compact and 
${\rm dist}_\bS:\bS\times \bS\to[0,\infty)$ 
is continuous, the distance minimum with respect to $q$ in the definition of $D$ exists.
It is strictly positive,
since by assumption $\cD_0 \vee \ldots  \vee \cD_{k_{\max}}=\{N\}$ and since 
$\Delta_{\{N\}}^E\cap \bS = \emptyset$.
\item 
The first inequalities in \eqref{dist:lob} and \eqref{dist:lob:k}
follow from the definition of ${\rm dist}_\bS$.
\item 
If the inequality $ \sum_{\ell=2}^{k_{\max}}{\rm dist}_\bS\big(Q(t_{\ell-1}),Q(t_\ell)\big)\ge 
D$ in \eqref{dist:lob:k} were false, then for {\em any} $q\in Q([t_{0}, t_{k_{\max}}])$
the maximum of the distances to all $\Delta_{\cD_\ell}\cap \bS$ ($\ell=0,\ldots,k_{\max}$) 
would be strictly smaller than $D$, leading to a contradiction.

As by \eqref{kmax} $k_{\max}\le n-1$, the second lower bound in \eqref{dist:lob:k}
follows.
\hfill $\Box$
\end{enumerate}
\begin{remark}[dependence of $D$ on the trajectory]\quad\label{rem:D}\\
The constant $D$ defined in \eqref{def:D} depends on the trajectory only via
the set partitions $\cD_\ell\in \cP(N)$, $\ell=0,\ldots,k_{\max}$. 
Since $k_{\max} \le n-1$ and $\cP(N)$ is finite, by 
taking the minimum of the $D(\cD_1,\ldots,\cD_{k_{\max}})$, 
there is a positive lower bound in \eqref{dist:lob} independent
of the choice of the trajectory. This modified $D>0$ then only depends on $d$, $n$ 
and the mass ratios.
\hfill $\Diamond$
\end{remark}
Now we inductively define $T_\ell := t_{k_{\max}}$ for initial time $T_{\ell-1}$ ($\ell\in \bN$)
with $T_0 := 0$ and set 
$\Delta_\ell := T_\ell- T_{\ell-1}\in (0,T^+ - T_{\ell-1})$.

We call the times $T_\ell$ {\bf chain-closing} (with respect to $T_{\ell-1}$).
Their index $\ell$ turns out to be relevant for the radial dynamics.
We introduce the following quantities: 
\begin{align*}
	G_\ell &:= \# \{ i \in \bN \mid 2\, i \leq \ell ,\ \Delta_{2i-1} \leq \Delta_{2i} \} \, , \\
	U_\ell &:= \# \{ i \in \bN \mid 2\, i \leq \ell ,\ \Delta_{2i-1} \geq \Delta_{2i} \} \, .
\end{align*}

\begin{theorem}[exponential lower bounds for system quantities] 
\label{thm:ExpLowerBound} \quad\\
The following statements are true, if the initial data satisfy $P_R(0)>0$:
\begin{enumerate}[1.]
\item 
We have for any $\ell \in \bN$:
\begin{align}
P_R(T_\ell) &\geq P_R(0) \cdot \left(1+D^2\right)^{U_\ell}, \label{eq:lowBoundPRTl} \\
R(T_\ell) &\geq R(0) \cdot \left(1+D^2\right)^{G_\ell}. \label{eq:lowBoundRTl}
\end{align}
\item 
If $\,\limsup_{\ell \rightarrow \infty} \Delta_{\ell}< \infty$, the radial momentum is 
bounded below exponentially in the number of chain-closing collisions, i.e.\ there is a 
constant $C_P>1$ depending only on $n$ and the particle masses such that for all but finitely many $\ell \in \bN$
\begin{align*}
P_R(T_\ell) \geq  A \, C_P^\ell \,,
\end{align*}
with $A>0$ independent of $\ell$ (but depending on $P_R(0)$, $R(0)$ and 
$\underset{\ell \rightarrow \infty}{\limsup} \Delta_{\ell}$).
\item 
For dimension $d=1$, if the particles are reflected during collision, i.e.\ the ordering of the
particles is never changed, then there is a constant $C_R>1$ depending only on $n$ and 
the particle masses such that for any $\ell \in \bN$
\[R(T_\ell) \geq C_R \, R(T_{\ell - 1}) \,.\]
\end{enumerate}
\end{theorem}
\begin{remarks}[significance and variants of Theorem \ref{thm:ExpLowerBound}]
\quad \\[-6mm]
\begin{enumerate}[1.]
\item 
Whereas the proof of Statement 3,\ uses a geometric argument,
for Statements 1.\ and 2.\ of Theorem~\ref{thm:ExpLowerBound} we use 
purely analytical considerations. 
Therefor, it is important, whether the time differences $\Delta_\ell$ are 
small or large and what can be said about the quotient of them. \\
Note that most of the statements can be proven analogously (with a smaller base), 
as long as the quotient $\frac{\Delta_{2i-1}}{\Delta_{2i}}$ is bounded below or above 
by some constant, for $G_{\ell}$ or $U_{\ell}$ suitably redefined.
\item 
\eqref{eq:lowBoundPRTl} and \eqref{eq:lowBoundRTl} essentially say that 
at least one of the quantities $P_R$ or $R$ is bounded below exponentially in the 
number of chain-closing collisions, as $G_{2\ell} + U_{2\ell}\ge \ell$.
Whether this applies to $G_{2\ell}$ or $U_{2\ell}$ may depend on $\ell$.
\item 
If we have a nondeterministic trajectory with finite escape time, the condition 
$\limsup_{\ell \rightarrow \infty} \Delta_{\ell} < \infty$ is automatically satisfied. 
\item 
For the growth of the radial momentum $P_R$ 
the most difficult case will be that of
times $\Delta_\ell$ between chain-closing collisions growing super-exponentially.
\item 
With some restriction on the possible combinatorics at collision times as in 
Statement 3.\ of Theorem~\ref{thm:ExpLowerBound} we can use some interesting 
geometric approaches to prove lower bounds for the growth of the radius. 
The ordering of the particles is, however, an important restriction.
\hfill $\Diamond$
\end{enumerate}
\end{remarks}

\noindent
\textbf{Proof of Theorem~\ref{thm:ExpLowerBound}:}\\[-6mm]
\begin{enumerate}[1.]
\item 
We start with some simple fact about the length of the curve on the sphere and the 
integral of $\|W\|_{\cM^{-1}}^2$: \\
As $\int_{T_{\ell-1}}^{T_\ell} \|\dot{Q}\|\,\mathrm{d}s 
= \mathrm{Length}(Q,[T_{\ell-1},T_\ell])\ge D$,
by the Cauchy-Schwarz inequality
\beq
\int_{T_{\ell-1}}^{T_\ell} \|W(s)\|_{\cM^{-1}}^2\,\mathrm{d}s \ge 
\frac{\mathrm{Length}(Q,[T_{\ell-1},T_\ell])^2} {\Delta_\ell}
\ge \frac{D^2}{\Delta_\ell}
\qquad(\ell\in \bN).
\Leq{Cauchy:Schwarz}
Now we will compare the growth of $R$ and $P_R$ during two subsequent chain-closing 
collisions. We will see that $R$ growths at least by a factor, if the fraction of time differences 
for the collisions is large, whereas $P_R$ will grow in the opposite case. 
\begin{align*}
R(T_{\ell}) &\geq R(T_{\ell -1}) + \Delta_\ell \, P_R(T_{\ell -1})  \\ 
& \geq R(T_{\ell -1}) + \Delta_\ell \, R(T_{\ell - 2}) \int_{T_{\ell-2}}^{T_{\ell-1}} 
\norm{W(s)}_{\cM^{-1}}^2 \,\mathrm{d}s \\ 
& \geq R(T_{\ell -2}) \,\left(1+D^2 \frac{\Delta_\ell}{\Delta_{\ell-1}} \right).
\end{align*}
Note that in the second line we used \eqref{PR:PR0} and in the last line we used 
\eqref{Cauchy:Schwarz} and the monotonicity of $R$. This proves \eqref{eq:lowBoundRTl} 
(note that $R$ is monotone, so for odd $\ell$ we can estimate 
$R(T_{\ell}) \geq R(T_{\ell-1})$).  \\
For the estimation of $P_R$ we use again \eqref{PR:PR0} but this time in another way:
\begin{align*}
P_R(T_{\ell}) &= P_R(T_{\ell -2})+\int_{T_{\ell -2}}^{T_{\ell}} \big(\textstyle R(T_{\ell-2}) +
\int_{T_{\ell -2}}^s P_R(\tau)\,d\tau\big)\,\|W(s)\|_{\cM^{-1}}^2\,\mathrm{d}s \geq \\& 
\geq P_R(T_{\ell -2})
+\int_{T_{\ell -1}}^{T_{\ell}} \Delta_{\ell -1} \cdot  P_R(T_{\ell-2}) \cdot \|W(s)\|_{\cM^{-1}}^2\,\mathrm{d}s \; + \\ & \qquad  \qquad 
+  R(T_{\ell-2}) \cdot \int_{T_{\ell -2}}^{T_{\ell}}\|W(s)\|_{\cM^{-1}}^2\,\mathrm{d}s \geq \\ 
& \geq P_R(T_{\ell-2}) \left(1+D^2 \frac{\Delta_{l-1}}{\Delta_{l}} \right)
+\left(\frac{1}{\Delta_{\ell-1}} + \frac{1}{\Delta_{\ell}}\right) D^2 R(T_{\ell -2}) \, .
\end{align*}
Using only the first summand, we have proven \eqref{eq:lowBoundPRTl}.
\item 
Under the assumption $\Delta_{\ell} \leq C$ for some value $C$ and all 
(but finitely many) $\ell \in \bN$, we can rewrite the inequality on $P_R(T_{\ell})$ as:
\begin{align*}
P_R(T_{\ell}) &\geq P_R(T_{\ell-2}) \left(1+D^2 \frac{\Delta_{l-1}}{\Delta_{l}} \right)
		+ \frac{1}{C} D^2 R(T_{\ell -2}) \geq \\ 
& \geq P_R(0) \left(1+D^2\right)^{U_\ell} + \frac{D^2}{C} R(0) \left(1+D^2\right)^{G_{\ell-2}}.
\end{align*}
But as $U_{\ell} + G_{\ell-2} \geq\lfloor \frac{\ell}{2}\rfloor-1$, at least one of 
the summands is larger than $\frac{\ell}{4}-1$. 
Choosing $C_R = \left(1+D^2\right)^{\frac{1}{4}}$ and 
$A= \frac{\min \{P_R(0), \frac{D^2}{C} R(0) \}}{1+D^2}$ implies the claim.
\item 
Considering the growth of the radius $R$, we know from Lemma \ref{lem:long}, 
that there is a $k\in \{1,\ldots,k_{\max}\}$ and an interval 
$[t_{k-1},t_k]\subseteq [T_{\ell-1},T_\ell]$ 
such that ${\rm dist}_\bS\big(Q(t_{k-1}),Q(t_k) \big) \,\ge\, \frac{D}{n}$.
We consider the tangent space 
\[U:=T_{q(t_{k-1})}\big(R(t_{k-1})\bS\big) \subseteq M\] 
through $q(t_{k-1})$ of the sphere of radius $R(t_{k-1})$, see Figure \ref{fig:Theorem3.15}. 

\begin{figure}[h]
\centerline{\includegraphics[width=60mm]{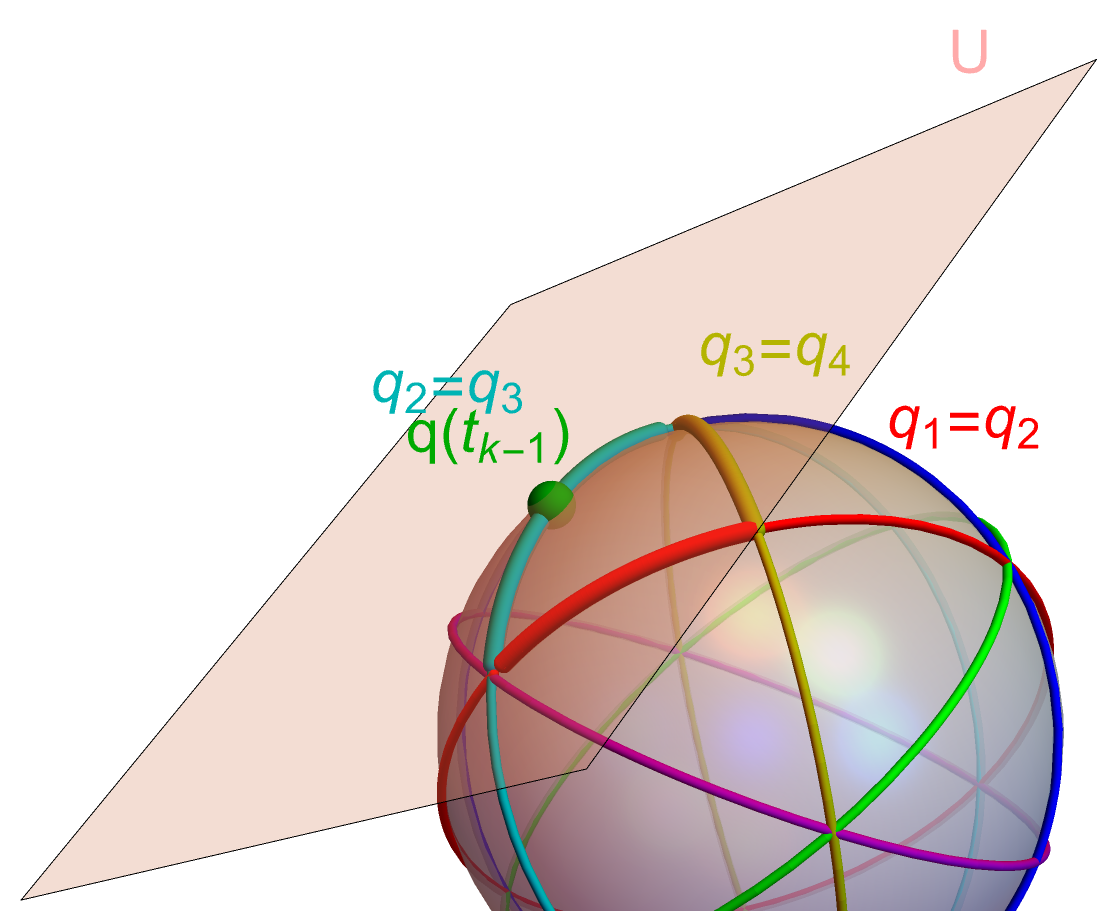}}
\caption{For $n=4$ particles in $d=1$ dimensions: the plane $U$ tangent to the sphere 
of radius $R(t_{k-1})$ at the point $q(t_{k-1})$ (shown in green), controlling the motion 
within the cone $\{q\in M \mid q_1\le q_2\le q_3\le q_4\}$.}
\label{fig:Theorem3.15}
\end{figure}

We claim that for $t\in (t_{k-1},T^+)$, $q(t)$ is not contained in the closed half-space 
$H(U)\subseteq M$ bounded by the affine hyperplane $U$ and containing 
$0\in M$. 
	
This is obvious for times $t > t_{k-1}$ if there is no intermediate collision,
since then in the interval $[t_{k-1},t]$ the motion $s\mapsto q(s)$ is on a straight 
line having the initial radial velocity $R'(t_{k-1}) > 0$.
	
For collisions with a hyperplane $\Delta_{i,j}$ at time $t$, the component 
\[\langle q'(t^\mp), q(t_{k-1})/\|q(t_{k-1})\|_\cM\rangle_\cM\] 
of velocity perpendicular to  
$U$ before/after collision is by assumption positive before collision. 
It increases during collision, since
both $q(t_{k-1})$ and $q'(t^+)$ belong to the half space bounded by 
$\Delta_{i,j}$ that contains the $(n-1)$--dimensional simplicial cone. 
Here we used that $d=1$.
	
In particular we have shown that $q(t_k)\not\in H(U)$.
Thus by \eqref{dist:lob:k},  with $C := \big(\!\cos(D/n)\big)^{-1} > 1$
\[\|q(t_k)\|_\cM\ge C\,\|q(t_{k-1})\|_\cM \, ,\]
which implies $R(T_\ell) \ge C \, R(T_{\ell-1})$ and inductively
$R(T_\ell) \ge C^\ell \, R(T_{0})$.
\hfill $\Box$\\[2mm]
\end{enumerate}
The relative radial velocity $P_R/R$ is not necessarily monotone increasing. Still:
\begin{lemma} \label{lem:log:R:derivative}
If the escape time $T^+$ is finite, then $\lim_{t\nearrow T^+}\frac{P_R(t)}{R(t)}=+\infty$.
\end{lemma}
\textbf{Proof:}\\[-6mm]
\begin{enumerate}[$\bullet$]
\item 
Since $\lim_{t\nearrow T^+}R(t)=\infty$, for $T^+<\infty$ the limes superior of
the relative radial velocity $P_R/R = \ln(R)'$ is infinite. 
\item 
This logarithmic derivative of $R$ fulfils the differential equation
\[ \textstyle
\frac{d}{dt} \frac{P_R}{R} = \|W\|_{\cM^{-1}}^2-\big(\!\frac{P_R}{R}\!\big)^2 
\ge -\big(\!\frac{P_R}{R}\!\big)^2.\]
For all $T\in [0,T^+)$ the initial value problem 
$x'=-x^2$, $x(T) = \big(\!\frac{P_R}{R}\!\big)(T)$
has the solution 
\[ \textstyle
x(t) = \frac{x(T)}{1+x(T)(t-T)} 
\ge \frac{x(T)}{1+x(T)(T^+-T)} 
\ge \eh \min \big(x(T),\frac{1}{T^+-T}\big) 
> 0 \]
in the interval $t\in [T,T^+)$. 
This implies $\frac{P_R}{R}(t) \ge  \eh  \min \big( \frac{P_R}{R}(T) , \frac{1}{T^+-T}\big)$.
\item 
For all $C>0$ there exists a $T\in [0,T^+)$ with $\frac{P_R}{R}(T)\ge 2C$ and 
$ \frac{1}{T^+-T}\ge 2C$. Thus $\frac{P_R}{R}(t) \ge C$ for all $t\in [T,T^+)$,
showing the claim.
\hfill $\Box$
\end{enumerate}

\subsection{Closedness of nondeterministic particle systems}

Definition~\ref{def:ND} of nondeterministic systems is concise, 
but the intuitive interpretation of particles
having collisions at which the momenta are changed (conserving total 
momentum but not necessarily kinetic energy) might be hidden. 

Additionally, this definition allows for accumulation points of the collision times in the 
interior of the domain of definition $[0, T^+)$. 
These might occur as some limiting behavior of trajectories with finitely many collisions 
in a given time interval when letting the number of collisions tend to infinity, e.g.\ 
in the model considered in \cite[Def.\ 3.2]{Qu} (but without allowing mass exchange). 

As in both definitions the information about the system is given by a continuous curve 
$q$, one natural notion of convergence is the convergence of the curves on any (or a 
given) compact subinterval of the domain of definition. 
\\
In this section we want to consider \textbf{two questions} of this kind:
\begin{enumerate}[1.]
\label{questions}
\item
Is the set of trajectories given by Definition~\ref{def:ND} closed with respect to 
convergence on compact intervals, i.e.\ is the limit curve again part of a
nondeterministic trajectory? 
This would show that nondeterministic systems share an important property of 
deterministic dynamical systems.
\item \label{question:2}
Can one uniformly approximate a nondeterministic trajectory $q$
with accumulation points of $\cT$ in the domain of definition $(T^-,T^+)$
by nondeterministic trajectories with only finitely many collisions in any 
compact interval? \\
Then one could use this approximation to prove properties of the trajectory~$q$,
using results obtained by combinatorial methods in \cite{Qu}.
\end{enumerate}
We will answer positively both questions, beginning with Question 1:
\begin{proposition}[Convergence of trajectories]\label{prop:conv:trajectories}
\quad \\
Let $(q^{(k)})_{k \in \bN}$ be a sequence of nondeterministic trajectories that converges 
uniformly on a compact interval $K$ to some map $q:K\to M$.
Then $q$ is part of a nondeterministic trajectory. 
\end{proposition}
\textbf{Proof:}\\
Denote with $\cT^{(k)}$ respectively $\cT$ the corresponding sets of $q^{(k)}$ and $q$ 
that were defined in Definition~\ref{def:ND} and analogously for $\CS^{(k)}$ and $\CS$. \\
We prove now the properties 2) and 3) of Def.\ \ref{def:ND} for $q$ 
to be part of a nondeterministic trajectory. 
We begin with considering a $\tau \in K$ and proving the differentiability of the map 
$q_{\CS(\tau)}^{E}$ in a small neighborhood of $\tau$. 
We treat the case $\tau \in \mathrm{int}(K)$, $\tau\in \partial K$ being similar.
By continuity of $q$ and the definition of $\CS(\tau)$ there is a
neighborhood $U=(a,b) \subseteq K$ of $\tau$ and an $\eps>0$ such that
\begin{align*}
\min_{\substack{i,j \in N, i \not\sim_{\CS(\tau)} j, \\ t \in U}} \norm{q_i(t) - q_j(t)} 
> \epsilon \, ,
\end{align*}
where $\sim_{\CS(\tau)}$ denotes the equivalence relation with equivalence classes 
$\CS(\tau)$. 
Due to the uniform convergence there exists an index $N_0 \in \bN$ such that for any $k \geq N_0$ we have
\begin{align*}
\min_{\substack{i,j \in N, i \not\sim_{\CS(\tau)} j, \\ t \in U}} \norm{q^{(k)}_i(t) - q^{(k)}_j(t)} 
> \frac{\epsilon}{2} \, .
\end{align*}
This implies that for $t \in U$ we always have $\CS(t) \preccurlyeq \CS(\tau)$ and 
even for indices $k$ sufficiently large we have $\CS^{(k)}(t) \preccurlyeq \CS(\tau)$. 
Note that this implies that $p_{\CS(\tau)}^{E, (k)}$ is constant on $U$, as we can project the 
locally constant functions $p_{\CS^{(k)}(\tau)}^{E, (k)}$ on this coarser partition (compare the 
proof of Lemma~\ref{lem:projection}). 
But then we get a simple expression for $q_{\CS(\tau)}^{E, (k)}$ in the interval $U=(a,b)$, 
as the motion of the external component is affine:
\begin{align*}
q_{\CS(\tau)}^{E, (k)}(t) = q_{\CS(\tau)}^{E, (k)}(a) + \frac{t-a}{b-a} \left(q_{\CS(\tau)}^{E, (k)}(b) - q_{\CS(\tau)}^{E, (k)}(a)\right).
\end{align*}
Now we can use the uniform convergence and obtain
\begin{align*}
q_{\CS(\tau)}^{E}(t) = q_{\CS(\tau)}^{E}(a) + \frac{t-a}{b-a} \left(q_{\CS(\tau)}^{E}(b) - q_{\CS(\tau)}^{E}(a)\right)\qquad (t\in U).
\end{align*}
This implies that $q_{\CS(\tau)}^{E}$ is affine in $U$ and hence especially differentiable. 
So Part~3) of Definition~\ref{def:ND} is satisfied. \\
What remains is to show Part 2), constancy of $p$ on any interval of $K\setminus \cT$. 
On such an interval $U \ni \tau$ we know that $p_{\CS(\tau)}^{I}$ is 0 and hence it is 
sufficient if $p_{\CS(\tau)}^{E}$ is constant. But this can be proven in the same way as above. 
\hfill $\Box$\\[4mm]
Now we consider the {\em second question} on Page \pageref{questions}, 
namely whether one can approximate any trajectory $q:I\to M$
with accumulation points of $\cT$
by trajectories without accumulation points in $I$. 
This will be answered in Prop.\ \ref{prop:ApproxFinitelyManyCollisions}.
	
We will focus on the one-dimensional case and consider first an isolated 
accumulation point of $\cT$. 
\begin{remark}[Simplifications for isolated accumulation points] \quad\\
If in a given compact time interval 
there is only one isolated accumulation point of ${\cal T}$ for the 
nondeterministic trajectory, we can use the following simplifications. 
\begin{enumerate}[1.]
\item 
By regarding the corresponding sub-clusters colliding at $T_{\mathrm{coll}}$,
we assume without loss of generality that this is a total collision 
$T_{\mathrm{coll}}$ (which is always isolated).
\item 
We can treat both sides of the time interval around $T_{\mathrm{coll}}$ separately,
starting with $[T_{\mathrm{coll}}-\epsilon, T_{\mathrm{coll}}]$. (The other case can be 
treated analogously by time reversibility of nondeterministic trajectories).\\
In this interval, we can enumerate all collision times in 
$[T_{\mathrm{coll}}-\epsilon, T_{\mathrm{coll}}) \cap \mathcal{T}$ 
in a monotone way (see Item 3.\ below): 
$t_1 < t_2 < \ldots$, with 
$\lim\limits_{k \rightarrow \infty} t_k = T_{\mathrm{coll}}$. 
\item 
Additionally, we can split up simultaneous collisions or collisions of more than 
two bodies in a sequence of binary collisions happening at the same physical time. 
This was treated in \cite[Section 3.1.2 and Section 3.1.5]{Qu}. 
Note that this is only possible as we do not demand conservation of energy.
\hfill $\Diamond$
\end{enumerate}
\end{remark}
By Theorem \ref{thm:finer}.3 we know that an accumulation point $T_{\mathrm{coll}}$
can be the limit of 
other accumulation points of collision times $\tau$ only if the latter have a  partition 
$\mathrm{CS}(\tau)$ of the particles that is strictly 
finer than $\mathrm{CS}(T_{\mathrm{coll}})$. \\
Hence the above analysis can be applied to the finest partitions first and 
then will yield the statement for arbitrary accumulation points. 
In such a case we must have a total collision of all particles 
of the same cluster at some time, as otherwise the diameter 
or the radial component of the momentum of the corresponding system 
would explode in backwards time direction.
	
We will now approximate the trajectory by ones that have only finitely many collisions 
within compact intervals.
As we can start this approximation arbitrarily close to the time of total collision, 
we will get a converging sequence of trajectories with only finitely many collisions.

To obtain such an approximation, we use a particular class of trajectories that 
we call {\em sticky}. 
\begin{defi}[sticky trajectories] \quad\label{def:ord:sticky} \\
A non-deterministic trajectory $\, q: I\to M$ of $\,n$ particles in $d$ dimensions
is \textbf{sticky}, if its set partition gets coarser in time: 
$\CS(t_1) \preccurlyeq \CS(t_2)$ for $t_1<t_2\in I$.
\end{defi}	
We also apply the notion of stickyness to restrictions of $q$ to subintervals of $I$.
But since in general one cannot uniquely extend a sticky trajectory backwards in
time, for our purposes it will be sufficient to consider the time interval $[0,\infty)$.
Then sticky trajectories are parameterised by their initial conditions in the 'sticky'
phase space 
\[ P_S := \{ (q,v) \in M\times M \mid v_\cC^I = 0 \mbox{ if } q\in \Xi_\cC \} \,  .\] 
In this sense they form a deterministic backbone of the
nondeterministic system:
\begin{lemma}[sticky trajectories] \quad\\ 
For all initial conditions $(q_0,v_0) \in P_S$ 
there exists a unique sticky trajectory $q: [0,\infty) \to M$ with 
$\big( q(0) , q'(0) \big) = (q_0,v_0)$.
\end{lemma}
\textbf{Proof:}
Consider the maximal interval $[0,t_1)$ of the ray $t\mapsto q(t) := q_0+v_0 t$ 
that is contained in $\Xi_{\CS(0)}$.
As the $\CS(0)$--internal velocity $(v_0)_{\CS(0)}^I$ equals $0$, 
either $t_1 = \infty$ or $q(t_1) \in \Xi_{\CS(t_1)}$ with 
$\CS(0) \prec \CS(t_1)$. In the latter case we set $v(t_1^+) := \Pi^E_{\CS(t_1)}(v(0))$, 
see \eqref{proj:E}, and proceed with
the new initial data $(q(t_1),v(t_1^+)) \in P_S$. 
As the rank of the partition decreases at each collision time 
\[ t_0 := 0 < t_1 < \ldots < t_k \, , \] 
one has $k \le n-1$,
and the inductive definition yields a sticky forward trajectory $q: [0,\infty) \to M$. 
\hfill $\Box$
\begin{remark}[sticky total collisions on a line]  
In the case $d=1$ of $n$ ordered ($q_1 \leq q_2 \leq \ldots  \leq q_n$) 
particles on a line, initial conditions 
$(q,v)\in P_S$ give rise to a sticky trajectory ending in a total collision, 
if $v_i > v_{i+1}$ for $1 \le i < n$. \\
This condition, however is not necessary. 
{\em E.g.}\ for $n=3$ it could be that $v_2 \le v_3$   
but we still get a total collision if $v_{\{1,2\}} > v_3$. 
\hfill $\Diamond$
\end{remark}
In all dimensions $d$, the cluster barycenter moment $J_\cC$, defined in 
\eqref{cluster:barycenter:moment}, allows to determine total collision time 
$T_{\mathrm{coll}}$ from the initial data of a sticky trajectory:
\begin{lemma}[total collision time for sticky trajectories] 
\label{lem:totalCollTimeStickyTrajectories} \quad\\ 
For a sticky trajectory $q$ with initial conditions $(q_0,v_0)\in P_S$ and 
$\cT= \{t_0,\ldots, t_k\}$,
that has a total collision at time $T_{\mathrm{coll}}:= t_k \in (0,\infty)$,  
\beq
T_{\mathrm{coll}} \ = \ 
S_{\mathrm{coll}} := 
- \tfrac{2J_\cC(0)} {J_\cC'(0)} \qquad \mbox{, with }\cC := \CS(t_{k-1}) \, .
\Leq{T:equal:S} 
\end{lemma}
\textbf{Proof:}
Before $T_{\mathrm{coll}}$ particles of different clusters of $\cC$ do not
collide. So the mean velocity $v_\cC = \Pi^E_\cC(v_0)$
is constant on $[0,T_{\mathrm{coll}})$,
the centers of mass of the clusters moving on the straight line 
$q_\cC(t) = (t - T_{\mathrm{coll}})v_\cC$.
The condition $T_{\mathrm{coll}} > 0$ implies that $v_\cC\neq 0$. So  
the quotient $S_{\mathrm{coll}}$ is well-defined and equals 
$\tfrac{\|v_\cC(0)\|_{\mathcal{M}}^2 T_{\mathrm{coll}}^2} {\|v_\cC(0)\|^2_{\mathcal{M}} T_{\mathrm{coll}}} 
= T_{\mathrm{coll}}$.\ 
\nolinebreak[4]\hfill $\Box$
\begin{remark}[sticky total collisions in dimension $d=1$]\quad\\ 
We enumerate the particles and clusters so that $q_i\le q_{i+1}$, 
and $\cC = \CS(t_{k-1}) = \{C_1,\ldots, C_R\}$ with $i<j$ if 
$q_i\in C_a$, $q_j\in C_b$ and $a<b$. The ordering $q_i\le q_{i+1}$ of the particles
do not change with time for a sticky trajectory.
\begin{enumerate}[1.]
\item 
Then for any set partition $\cD$ with  $\cC \preccurlyeq \cD \prec \cC_{\max} = \{N\}$, 
whose atoms are unions of neighbouring atoms of $\cC$, the mean cluster 
velocity $v_\cD$ is non-zero and constant, too on $[0,T_{\mathrm{coll}})$,
and the analog of \eqref{T:equal:S} for $\cD$ is valid. 
\item 
In particular, there exist such set partitions $\cD$ of rank two. 
For $|\cC|-1$ specific values of $r$ they are of the form 
\beq \cD_r := \{D_1,D_2\}\mbox{ , with } 
D_1 = \{1,\ldots, r\} \mbox{ and } D_2 = \{r+1,\ldots, n\} \, . 
\Leq{D:r:independent} 
\item 
$\{\cD_1, \ldots, \cD_{n-1} \} \subseteq \cP(N)$ is an antichain for the poset 
$\cP(N) \backslash \{\cC_{\max}\}$.\\
For {\em all} $1 \le r \le n-1$ we have $q_{D_1} < 0 < q_{D_2}$ and 
$v_{D_2} < 0 < v_{D_1}$, so that the denominator in the analog
$- 2 J_{\cD_r}(0) / J_{\cD_r}'(0)$ of \eqref{T:equal:S} is positive.
\item 
By Item 1., if  $\cD_r \succcurlyeq \cC$, $v_{\cD_r}$ is non-zero and constant, with
$v_{\cD_r} (t) = \Pi^E_\cC(v_0)$. 
Then the two cluster centers are moving according to 
$q_{\cD_r} (t) = (t - T_{\mathrm{coll}}) v_{\cD_r}$.
Otherwise, if $\cD_r \not\succcurlyeq \cC$, then 
$v_{\cD_r}$ is still piecewise
constant between consecutive collision times $t_\ell$. But there can be
an index $\ell$ (necessarily unique) with 
$\cD_r \succcurlyeq \CS(t_{\ell -1})$ and $\cD_r \not\succcurlyeq \CS(t_{\ell})$.
If so, 
\begin{enumerate}[$\bullet$]
\item 
the collisions at times $t_1,\ldots, t_{\ell -1}$ occur within $D_1$ or $D_2$ 
and thus leave the cluster velocity $v_{\cD_r}$ unchanged.
\item 
However, at time $t_\ell$, and possibly at later collision times $t_{\ell'}$, 
$0 < v_{D_1}^+ < v_{D_1}^-$ (and $v_{D_2}^+ > v_{D_2}^- > 0$), because otherwise
the $r$-th and $(r+1)-$th particle would not have met.
So the function
\beq 
S _r : [0,T_{\mathrm{coll}} ) \to \bR \quad ,\quad 
S_r(t) = -2 J_{\cD_r}(t)/J_{\cD_r}'(t) 
\Leq{S:r}
jumps up at $t_\ell$
\item
Between collisions $S_r$ is piecewise affine, with slope $-1$:
The center of mass is at 0, so $m_2 q_2 = - m_1 q_1$ and 
$m_2 v_2 = - m_1 v_1$ for $m_i := m_{D_i}$ etc.\
with $q_i^{(\ell)} := q_i(t_\ell)$ and $v_i^{(\ell)}:= v_i(t_\ell^+)$, and  this entails for 
$t\in(t_\ell,t_{\ell+1})$: $S_r(t) =$
\beq 
\tfrac{m_1 q_1^2(t) + m_2q_2^2(t)} {-m_1 q_1(t) v_1(t) - m_2 q_2(t) v_2(t)}
= \tfrac{q_1(t) - q_2(t)}
{-v_1(t) + v_2(t)} = \tfrac{q_1^{(\ell)} + v_1^{(\ell)} t - q_2^{(\ell)} - v_2^{(\ell)} t}
{-v_1^{(\ell)} + v_2^{(\ell)}}
=  \tfrac{q_1^{(\ell)}}{-v_1^{(\ell)}} - t \, .
\Leq{slope:minus:one}
\item
Since in the time interval $(t_{k-1},t_k)$ all the clusters in
$\cC$ move with constant velocities towards total collision at time $t_k $,
then $S_r(t) = T_{\mathrm{coll}} - t$ for {\em all} $r$. 
Since only those $S_r$ with  $\cD_r \not\succcurlyeq \cC$ are discontinuous, 
having jumps upwards, this shows that
\beq
S_{\mathrm{coll}} 
= \max \big\{ S_r(0) \, \big| \; 1\le r \le n-1 \big\} \, .
\Leq{S:coll:m}
\end{enumerate}
\end{enumerate}
The identity \eqref{S:coll:m} for total collision time 
$T_{\mathrm{coll}} = S_{\mathrm{coll}}$ of sticky trajectories 
is more useful than the original definition \eqref{T:equal:S} of $S_{\mathrm{coll}} $, 
since it uses the  family of set partitions $\cD_r$ that, unlike $\cC$, 
is independent of the choice of initial conditions in the phase space $P_S$. 
As it will turn out below, \eqref{S:coll:m} will also be a lower bound for the 
total collision time $T_{\mathrm{coll}}$ of general nondeterministic trajectories.
\hfill $\Diamond$
\end{remark}
When we now analyse nondeterministic trajectories $q: I\to M$ in dimension $d=1$, 
experiencing a total collision, we have to take account of the fact, that 
the ordering of the particle positions $q_i(t) \in \bR$  can change at the collision times
$t_k\in \cT \cap I$. As a consequence of this, the $S_r$ from \eqref{S:r} 
can jump {\em down} at $t_k$.
 \\
We restrict to the time interval $I = [0, T_{\mathrm{coll}}]$, with total collision time 
$T_{\mathrm{coll}} > 0$ and now  use Eq.\ \eqref{S:coll:m}, 
valid for sticky trajectories, as the {\em definition} of $S_{\mathrm{coll}}$.
We generalise this to time $t$, with the deterministic estimate 
\beq
S_{\mathrm{coll}} : I\to \bR \quad, \quad  
S_{\mathrm{coll}}(t) = \max \big\{ S_r(t) \, \big| \; 1\le r \le n-1 \big\} 
\Leq{S:coll:new}
for the time left to total collision. \\ 
The question arises whether that maximum can jump down at times $t_k$ for a 
nondeterministic trajectory, like the $S_r$ can do.

For the moment we assume that the only accumulation point of $\cT\cap I$ is 
$T_{\mathrm{coll}}$, so that we can enumerate the collision times, 
with $t_k< t_{k+1} \in [0, T_{\mathrm{coll}})$.
\begin{lemma}[Lower bound for the time to a total collision] 
\label{lem:LowBoundTimeTotalCollision} \quad \\
With the above assumptions, for nondeterministic trajectories in dimension $d=1$, 
\\[-6mm] 
\begin{enumerate}[1.]
\item 
At all collision times $t_k$, one has 
$S_{\mathrm{coll}}(t_k^+) \ge S_{\mathrm{coll}}(t_k^-)$.
\item  
Total collision time exceeds its deterministic estimate:
$T_{\mathrm{coll}} \ge S_{\mathrm{coll}}(0)$.
\end{enumerate}
\end{lemma}
\textbf{Proof:}
\begin{enumerate}[1.]
\item 
There can only be a jump of $S_r$ at time $t_k$, if at this collision
particles from $D_{r,1}$ and $D_{r,2}$ collide. Collisions internal to one of these two
clusters do not influence the motion of their centers of mass.

We assume without loss of generality that
$q_1(t_k) \le q_2(t_k) \le \ldots \le q_n(t_k)$. 
\beq
R(t) 
:= \min \big\{r \in\{1,\ldots,n-1\} \mid S_r(t) = S_{\mathrm{coll}}(t) \big\} 
\mbox{ and }R_k:=R(t_k^-)
\Leq{R:k} 
is the smallest index of a maximiser. 
Note that $R(t)=R_k$ for $t \in (t_{k-1},t_k)$.
\begin{enumerate}[$\bullet$]
\item 
If $R_k = 1$, then $q_{D_{r,1}} < 0$ is continuous at $t_k$,
but $0 < v_{D_{r,1}} (t_k^+) \le v_{D_{r,1}} (t_k^-)$. So 
$S_{\mathrm{coll}}(t_k^+) \ge S_{\mathrm{coll}}(t_k^-)$
in \eqref{S:coll:new}.
\item 
$R_k = n-1$ leads to the same inequality, by considering $-q$ instead of
$q$.
\item 
So we are left with $2\le R_k \le n-2$. We claim that then there is no collision
at $t_k$ between particles $R_k$ and $R_k+1$, 
so that $S_{R}(t_k^+)=S_{R}(t_k^-)$.

To prove this, we consider for an arbitrary index $r\in \{2,\ldots,n-2\}$ the sign of 
$S_{r+1}(t_k^-) - S_{r}(t_k^-)$. To simplify notation, we omit time and write 
$D_r$ for the first cluster in $\cD_r = \{D_{r,1}, D_{r,2}\}$. Then
\begin{align*}
S_{r+1}-S_{r} & = \tfrac{m_{D_r} q_{D_r}} {m_{D_r} v_{D_r}} -
\tfrac{m_{D_r} q_{D_r} + m_{r+1}q_{r+1}} {m_{D_r} v_{D_r}+ m_{r+1}v_{r+1}}\\
&= \tfrac{(m_{D_r} q_{D_r})(m_{D_r} v_{D_r}+ m_{r+1}v_{r+1}) 
- (m_{D_r} q_{D_r} + m_{r+1}q_{r+1})(m_{D_r} v_{D_r})}
{(m_{D_r} v_{D_r})(m_{D_r} v_{D_r}+ m_{r+1}v_{r+1})} = \tfrac{N}{D} \, .
\end{align*}
The denominator $D$ is positive, since $r+1 < n$. So the sign we want to control
is the sign of the numerator $N$. But, masses being positive, this is the sign of
\beq
\tfrac{N}{m_{r+1}} = m_{D_r}q_{D_r} v_{r+1} -  m_{D_r}v_{D_r }q_{r+1} \, .
\Leq{sign:determining}
As we assumed maximality of $S_R$ {\em and} minimality of $R$, 
$S_{R+1}-S_R \le 0$.
For $r:=R$ we thus get 
\beq
0 \ge m_{D_R}q_{D_R} v_{R+1} -  m_{D_R}v_{D_R }q_{R+1} \, .
\Leq{SR+1-SR}
But if the particles $R$ and $R+1$ collide, $q_{R}(t_k) = q_{R+1}(t_k)$.
For them to collide, we must have $v_{R}(t_k^-) \ge v_{R+1}(t_k^-)$. Inserting
this into \eqref{SR+1-SR} gives
\[0 \ge  m_{D_R} q_{D_R} v_{R} -  m_{D_R} v_{D_R} q_{R} = 
m_{D_{R-1}} q_{D_{R-1}} v_{R} -  m_{D_{R-1}} v_{D_{R-1}} q_{R} \, .\]
However, this equals \eqref{sign:determining} for $r:=R-1$.
So we have shown that $S_{R-1} \ge S_R$, contradicting maximality of $S_R$
and/or minimality of $R$.\\
But without a cluster-external collision at time $t_k$ for $\cD_R$, we have 
$S_R(t_k^+)= S_R(t_k^-)$, now for all $R \in\{1,\ldots,n-1\}$.
The mechanism that leads to positive jumps 
($S_{\mathrm{coll}}(t_k^+) > S_{\mathrm{coll}}(t_k^-)$)
 is a collision between particles
$r$ and $r+1$, resulting in a positive jump of a formerly {\em non-maximal} $S_r$. 
\end{enumerate}
\item 
We now show that $T_{\mathrm{coll}} \ge S_{\mathrm{coll}}(0)$. 
To this aim we modify Def.\ \eqref{R:k} of $R_k$:
\beq
\widetilde{R}(t)  
:= \min \big\{r \in\{1,\ldots,n-1\} \mid S_r(t) \ge S_{\mathrm{coll}}(0) - t \big\}
\mbox{ and } \widetilde{R}_k := \widetilde{R} (t_k^-) \, .
\Leq{R:k:hat} 
As for $R$, $\widetilde{R}(t) = \widetilde{R}_k$ in the interval $(t_{k-1},t_k)$.\\
For all $k$ and  $t\in (t_k , t_{k+1})$, one has
$S_{\mathrm{coll}}(t) = S_{\mathrm{coll}}(t_k^+) + t_k - t$, 
as shown in \eqref{slope:minus:one}. So 
$S_{\mathrm{coll}}(t_{k+1}^-) = S_{\mathrm{coll}}(t_k^+) + t_k - t_{k+1}$.
As $S_{\mathrm{coll}}(t_k^+) \ge S_{\mathrm{coll}}(t_k^-)$ (see Item 1), it 
follows that $S_{\mathrm{coll}}(t_k^-) \ge S_{\mathrm{coll}}(0) - t_k$. So
the set in \eqref{R:k:hat} is non-empty, 
$\widetilde{R}_k\le R_k$ and $S_{\widetilde{R}_k} (t_k^-) \le S_{R_k}(t_k^-)$. 
We have 
\beq
q_{\widetilde{R}_{k+1}} (t_{k}) \le q_{\widetilde{R}_k} (t_k) \, ,
\Leq{q:falling}
since a collision at time $t_k$ 
\begin{enumerate}[$\bullet$]
\item 
between particles within $D_{\widetilde{R}}$ 
leaves $S_{\widetilde{R}}$ unchanged so that 
$\widetilde{R}_{k+1} \le \widetilde{R}_k$,
\item 
between particles of indices larger than $\widetilde{R}_k$ leaves 
$S_{\widetilde{R}}$ unchanged, too, thus $\widetilde{R}_{k+1} = \widetilde{R}_k$,
\item 
between particles $\widetilde{R}_k$ and $\widetilde{R}_k + 1$ implies 
$q_{\widetilde{R}_k}(t_k) = q_{\widetilde{R}_{k+1}}(t_k)$.
But in this case we have 
$S_{\widetilde{R}_k + 1}(t_k^{-}) \geq S_{\widetilde{R}_k}(t_k^{-})$, as otherwise we 
could argue as above and conclude 
$S_{\widetilde{R}_k}(t_k^{-}) \le S_{\widetilde{R}_{k}-1}(t_k^{-})$. 
This would be a contradiction to the minimality of $\widetilde{R}_k$. \\
So we have $S_{\widetilde{R}_k + 1}(t_k^{+}) \geq S_{\widetilde{R}_k + 1}(t_k^{-})$ 
and this implies $\widetilde{R}_{k+1} \leq \widetilde{R}_k + 1$ and hence the 
statement.
\end{enumerate}
This has the consequence that the piecewise continuous function 
\[ f : \big[0,S_{\mathrm{coll}}(0) \big) \to \bR \quad , \quad 
f(t) = \frac{-q_{\widetilde{R}(t)}} {S_{\mathrm{coll}}(0) - t}\]
has no downward jumps at its potential points $t_k$ of discontinuity.
$f(t)$ is the slope of the line through the points $(t,-q_{\widetilde{R}(t)})$ and 
$(S_{\mathrm{coll}}(0),0)$. 
For non-collision times 
$t\in \big[0,S_{\mathrm{coll}}(0) \big) \backslash {\cal T}$, $f$ is differentiable with
\[f'(t) = - \frac{ q_{\widetilde{R}}(t) + (S_{\mathrm{coll}}(0) - t)  v_{\widetilde{R}}(t)}
{(S_{\mathrm{coll}}(0) - t)^2} \, .\]
We claim that $f'(t)>0$. The denominator being positive, this follows from
\beq 
m_{\widetilde{R}}(t) q_{\widetilde{R}}(t) + 
(S_{\mathrm{coll}}(0) - t) \, m_{\widetilde{R}}(t) v_{\widetilde{R}}(t) < 0 \, .
\Leq{numerator:derivative}
The left hand side of \eqref{numerator:derivative} equals
\begin{align} 
&m_{D_{\widetilde{R}}}(t) q_{D_{\widetilde{R}}}(t) - 
m_{D_{\widetilde{R}-1}}(t) q_{D_{\widetilde{R}-1}}(t) \nonumber\\ 
&+
(S_{\mathrm{coll}}(0) - t) \, \big( m_{D_{\widetilde{R}}}(t) v_{D_{\widetilde{R}}}(t) - 
m_{D_{\widetilde{R}-1}}(t) v_{D_{\widetilde{R}-1}}(t) \big) \nonumber \\
=&\ m_{D_{\widetilde{R}}}(t) \big(q_{D_{\widetilde{R}}}(t) + 
(S_{\mathrm{coll}}(0) - t) v_{D_{\widetilde{R}}}(t) \big) 
\label{f:prime:1}\\ 
&- m_{D_{\widetilde{R}-1}}(t) \big(q_{D_{\widetilde{R}-1}}(t) + 
(S_{\mathrm{coll}}(0) - t) v_{D_{\widetilde{R}-1}}(t) \big) \, .
\label{f:prime:2}
\end{align}
By definition of $\widetilde{R}(t)$, \eqref{f:prime:1} is non-positive 
and \eqref{f:prime:2} is negative, proving \eqref{numerator:derivative}.
So $f$ is strictly increasing.

Now if we had $S_{\mathrm{coll}}(0) > T_{\mathrm{coll}}$, then 
$f(T_{\mathrm{coll}}) = 0$. 
If we can assume that $q_{\widetilde{R}}(0) \le 0$
so that $f(0) \ge 0$, then we have produced a contradiction.\\
If instead $q_{\widetilde{R}}(0) > 0$, then by 
Remark \ref{rem:q:cS:0}.4,\ $\widehat{q} := -q$ is
a nondeterministic trajectory, with $\widehat{q}_r = -q_{n-r}$, and thus $\widehat{q}_r(0) < 0$
for all $r\le n-R(0)$. If we denote its quantities by a hat, then 
$\widehat{T}_{\mathrm{coll}} = T_{\mathrm{coll}}$, $\widehat{S}_r = S_{n-r}$, and thus
$\widehat{S}_{\mathrm{coll}} = S_{\mathrm{coll}}$. As  $\widetilde{R}(0) = R(0)$,
$\widehat{S}_{n-\widetilde{R}(0)} = \widehat{S}_{\mathrm{coll}}(0)$  and  
$\widetilde{\widehat{R}}(0) = \widehat{R}(0) \le n-R(0)$. 
So $\widehat{q}_{\widetilde{\widehat{R}}}(0) < 0$ as desired.
\hfill$\Box$
\end{enumerate}
With this preparation we can positively answer Question 2 on Page \pageref{question:2}:

\begin{proposition}[Approximations with finitely many collisions] 
\label{prop:ApproxFinitelyManyCollisions} \quad \\
For every nondeterministic (forward) trajectory $q: [0, T) \rightarrow M$ and every 
$\eps>0$ there exists a nondeterministic trajectory 
$q^{(\eps)}: [0, T) \rightarrow M$ such that:
\begin{enumerate}[$\bullet$]
\item 
For any compact interval $K \subseteq [0, T)$ we have
\[ \lim\limits_{\eps \downarrow 0} \sup_{t \in K} 
\left\| q^{(\eps)}(t) - q(t) \right\| = 0 \, . \]
\item 
Each $q^{(\eps)}$ has only finitely many collision in any compact interval 
$K \subseteq [0, T)$.
\end{enumerate}
\end{proposition}
\textbf{Proof:}\\
Let $\eps>0$ be arbitrary. We construct the approximating trajectory as follows: 
We start with the collision times $t$ of finest partitions $\cC$ that are 
accumulation points of collision times in the interval $[0, T-\eps]$ 
(after time $T-\eps$ we continue the trajectory in an arbitrary way, e.g.\ the particles 
do not exchange any momenta anymore). 
As these collision times are isolated by Theorem~\ref{thm:finer}, there is an interval 
around each $t$, say $[t-\delta_t, t+\delta_t]$, in which only particles from the same 
clusters of $\cC$ collide. 
If there are infinitely many collisions within one cluster, these particles must contract 
to their common center of mass. 

Now we can apply Lemma~\ref{lem:LowBoundTimeTotalCollision} and 
Lemma~\ref{lem:totalCollTimeStickyTrajectories} just for the motion of the particles of 
this subsystem having a total collision at their relative center of mass. 
Without loss of generality (otherwise take a smaller $\delta_t$) we assume that 
$J(t\pm \delta_t) \leq \operatorname{Const} \cdot \eps^2$, 
where the constant compensates the norm equivalence of the norms $\norm{q}$ and 
the norm with respect to the mass metric $\norm{q}_{\mathcal{M}}^2 = 2 J$, 
and some other factors. 
We approximate the trajectory by the sticky trajectory from $t-\delta_t$ up to time $t$ 
and from time $t+\delta_t$ to time $t$ in backward-time direction. 
The above lemmas guarantee that the sticky trajectory has a time of total collision which is at 
most $t$, hence this approximation is well-defined and yields a nondeterministic trajectory. 
This depends on $\eps$ and can guarantee that the distance between our first 
approximation $q^{(\eps), 1}$ and $q$ is smaller than $\frac{\eps}{2}$ on each of these 
intervals (the distance between the trajectories can be estimated in terms of the 
difference of the internal coordinates which is small as the total internal moment of inertia 
is small on this interval due to convexity). \\
Next we can take all collision times from the next coarser partitions for $q^{(\eps), 1}$. 
These are now isolated accumulation points of collisions (due to the first step) and hence 
we can repeat the above procedure and approximate this trajectory by $q^{(\eps), 2}$ 
that has a distance of at most $\frac{\eps}{4}$ to $q^{(\eps), 1}$ at any time. 
We repeat this procedure a finite number of times (for each coarser and coarser partition) 
until we finally reach the desired $q^{(\eps)}$. 
\\
The convergence of the trajectories is clear as by construction we even have
\[ \sup_{t \in [0, T-\eps]} \left\| q^{(\eps)}(t) - q(t) \right\| \leq \eps \, . \]
That $q^{(\eps)}$ only has finitely many collisions is also true by construction. 
\hfill $\Box$
\begin{corollary}[Exponential lower bounds for system quantities] \quad \\
For an expanding nondeterministic system with $J'(0)>0$ the moment of inertia and the 
kinetic energy are exponentially lower bounded in the number of chain-closing collisions.
\end{corollary}
\textbf{Proof:}
\\
This follows from the corresponding statements \cite[Prop.\ 3.51]{Qu}
(concerning the kinetic energy) and 
\cite[Prop.\ 3.61]{Qu} (concerning the moment of inertia), 
for systems that only have an accumulation point at time $T$, using the above 
approximation from Proposition~\ref{prop:ApproxFinitelyManyCollisions}.
Note that the number of chain-closing collisions of the 
approximating systems differs from the original system at any time by at most 1. 
The approximation can have a chain-closing collision that happens during 
a total collision of a subsystem of particles earlier or later than the original system, 
but at least at the time of the total collision of the subsystem the number of 
chain-closing collisions match again.
\hfill $\Box$
%
\section{Lagrangian relations for the non-deterministic system}\label{sec:Lagrangian}
%
A {\bf Lagrangian relation} is a Lagrangian submanifold 
of the symplectic product manifold 
$(X\times Y,\omega_\ominus:= \pi_X^*\omega_X-\pi_Y^*\omega_Y)$  
of symplectic manifolds $(X,\omega_X)$ and $(Y,\omega_Y)$.
An example is the graph of a symplectomorphism $X\to Y$.
Nondeterministic billiards do {\em not} lead to such graphs, even with
energy conservation.
\begin{example}[Lagrangian relation for a nondeterministic billiard] \label{ex:coll}\quad\\
The cotangent bundle, $X=Y:= T^*\bS^{d-1}$ of the sphere is naturally symplectic. 
The submanifold $L := \{(0,x;0,y)\in X\times Y \mid x,y \in \bS^{d-1}\}$ is Lagrangian
but not a graph. It could model the reflection of an otherwise
free particle in configuration space $\bR^{d}$ by the origin, with 
incoming data $(p_x,x)\in T^*\bR^{d}$, 
outgoing data \linebreak $(p_y,y)\in T^*\bR^{d}$ and energy conservation
(say, $\|p_x\| = \|p_y\| = 1$).\hfill$\Diamond$
\end{example}
The last example has been generalised in \cite{FKM}  to repeated reflections of
a point particle by finitely many linear subspaces of configuration space. 
Unlike in the present setting, 
the scattering events were assumed to preserve kinetic energy.
In this section we analyse multiple scattering without energy preservation.
We show that by prescribing finite sequences of scattering subspaces and
of kinetic energies a Lagrangian relation is defined.
%
\subsection{Scattering by a single subspace}
%

Our general goal is to use the non-deterministic system in the analysis of the
set ${\rm NC}$ of deterministic non-collision singularities.
However, it is not obvious how to compare the  two
dynamics in phase space~$P$, 
\begin{enumerate}[$\bullet$]
\item 
since, unlike the total energy of the Hamiltonian
dynamics, the non-de\-ter\-min\-is\-tic total (kinetic) energy can depend on time,
\item 
and since the deterministic dynamics is complicated at near-collisions.
\end{enumerate}
Since we are interested in measure-theoretic statements, we partially ignore these
difficulties by using Poincar\'e surfaces erected over submanifolds of $M$ 
near $\Delta$, much like in Example \ref{ex:coll}.\footnote{In an  
approach that is more intrinsic to the non-deterministic system, 
one could use instead the blow-up of $\Delta$, like in \cite{KM}, 
based on \cite{Me} of {\sc Melrose}.}

We begin by considering for $\cC\in \cP_\Delta(N) = \cP(N)\setminus\{\cC_{\min}\}$
the linear subspace $\Delta^E_\cC\subseteq M$ of positive codimension $d(n-|\cC|)$,
see \eqref{dim:MC}. 
With the sphere $\bS^I_\cC = \big(J^I_\cC\big)^{-1}(1/2)$
we consider in the $D = \dim (M) = (n-1)d$--dimensional
center of mass configuration space 
the codimension one cylinder 
\beq 
Z_\cC := \bS^I_\cC \times \Delta^E_\cC \subseteq M \quad\mbox{with }
\dim(Z_\cC) = D-1 \, .
\Leq{Z:C}
Equipped with  its natural symplectic form
$\omega_\cC$, its cotangent bundle 
\[P_\cC := T^*Z_\cC\]
is a $2(D-1)$--dimensional symplectic manifold. Like in Example \ref{ex:coll},
the $P_\cC$  can serve as a Poincar\'e surfaces (actually, in \cite{FK2}, variants 
were used to show that collisions are of measure zero for $-r^{-\alpha}$ pair 
potentials if $0<\alpha<2$). For the collision set $\Delta$ from \eqref{coll:set}, 
the potentials considered are of the form
\[ V: M\backslash\Delta \to \bR \quad,\quad 
V(q ) = \sum_{i<j} \tfrac{-Z_{i,j}}{\|q_i-q_j||^\alpha}\, .\]
It is important to see that there are two perspectives regarding $P_\cC$:
\begin{enumerate}[1.]
\item 
For the {\em deterministic} flow, the total energy $E$ of the Hamiltonian
\[ \textstyle
H \in C^\omega \big(T^*(M\backslash \Delta) , \bR\big) \qmbox{,} H(p,q) = K(p)+V(q)
\]
is fixed.
Then for 
${\cal H}_{E,\cC} := \{ (p,q) \in  H^{-1}(E) \mid q \in  Z_\cC\}$ 
the flow is transversal to
\[{\cal H}_{E,\cC}^\pm := 
\{(p,q)\in  {\cal H}_{E,\cC} \mid\pm \langle p^I_\cC,q^I_\cC\rangle>0 \} \, , \]
leaving respectively entering the cylinder $Z_\cC$.
Let $N: Z_\cC\to T_{Z_\cC} M$ be the (outward) unit normal vector field.
Then ${\cal H}_{E,\cC}^\pm$ both project diffeomorphically to their common image
in $P_{\cC}$, via the maps
\[n^\pm: {\cal H}_{E,\cC}^\pm \to P_{\cC} \qmbox{,}
(p,q)\mapsto \big( p - p(N(q)) N^{\flat}(q), q \big) \, .\]
The corestrictions of $n^\pm$ are even symplectomorphisms, see \cite[Thm.\ C]{FK1}.\\
In this sense ${\cal P}_\cC$ can serve as a Poincar\'e surface in $ \hSE$,
\item 
Regarding the {\em non-deterministic} dynamics, we noted in Remark \ref{rem:q:cS:0}.5 that
the trajectories come in one-parameter families, with initial (kinetic) energy as the 
parameter.  This comes in handy, since energy is not conserved along a
trajectory. The submanifolds
\[ Y_\cC := \{(p,q)\in P \mid q\in Z_\cC, p^I_\cC\neq 0\} \mbox{ and }
Y_\cC^\pm := \{(p,q)\in Y_\cC \mid \pm \langle q^I_\cC,p^I_\cC \rangle > 0\}\]
are of codimension one, and thus the restriction of the symplectic form $\omega$ on 
$Y_\cC$ has one-dimensional
kernels 
\beq
\mathrm{ker}_y(\omega):= \{v\in T_y Y_\cC \mid 
\forall\, w\in T_y Y_\cC: \omega_y(v,w) = 0\}\qquad (y\in Y_\cC).
\Leq{symp:kernel} 
As $Y_\cC$ is open in the level set of the regular value 
$1/2$ for $J^I_\cC$,
the characteristic foliation of $Y_\cC$ given by \eqref{symp:kernel}
is by  the orbits of the Hamiltonian flow 
\[\bR \times Y_\cC \to Y_\cC\qmbox{,}
(t;p,q) \mapsto \big(p-\cM \, q^I_\cC \, t,q \big)\]
of $J^I_\cC$ on $Y_\cC$.
So the leaf space $Y_\cC/\!\sim$ is naturally symplectomorphic to $\cP_\cC$,
which therefore can be used as a Poincar\'e surface, like in Case 1.
\end{enumerate}

Our first use of the Poincar\'e surfaces $\cP_\cC$ is to study scattering 
of a particle with initial energy $E_-$ by 
$\Delta^E_\cC$, changing its energy to $E_+$.
On $P_\cC\times P_\cC$ with projections 
$\pi_\pm:P_\cC\times P_\cC \to P_\cC$ to the right respectively
left factor, the two-form 
$\omega_\ominus := \pi_-^*(\omega_\cC)- \pi_+^*(\omega_\cC)$
is symplectic.

\begin{lemma}[Lagrangian relation]  \label{lem:Lag:rel}
For all $\cC\in \cP_\Delta(N)$ and $E_\pm>0$ the set
\begin{align}
L_\cC &\equiv L_{\cC;E_-,E_+} := 
\Big\{ (p_-,q_-; p_+ ,q_+) \in P_\cC\times P_\cC \,\Big|\, 
(p_\pm)^I_\cC = 0, (p_+)^E_\cC = (p_-)^E_\cC \, , \NN \\
&\hspace*{-4mm}
{\textstyle K^E_\cC(p_\pm) < \min(E_-,E_+),\ 
(q_+)^E_\cC - \frac{(\cM^{-1} p_+)^E_\cC}{\sqrt{2(E_+ - K^E_\cC(p_+))}} = 
(q_-)^E_\cC + \frac{(\cM^{-1} p_-)^E_\cC}{\sqrt{2(E_- - K^E_\cC(p_-))}} \Big\} }
\label{def:L:CEE}
\end{align}
is a Lagrangian manifold in $(P_\cC\times P_\cC, \omega_\ominus)$. 
\end{lemma}
\begin{remark}[significance of definition \eqref{def:L:CEE}]
\quad\label{rem:flight:times}\\[-6mm]

\begin{enumerate}[1.]
\item 
As the $(p_\pm)^I_\cC$ are the components of momenta $p_\pm$ 
(co)--tangential to the sphere $\bS^I_\cC$ in \eqref{Z:C}, 
the conditions $(p_\pm)^I_\cC=0$ are equivalent to collision with the subspace 
$\Delta_\cC$, if the radial momentum components do not vanish (see 3.).
\item
The condition $(p_+)^E_\cC = (p_-)^E_\cC$ means preservation of total momentum.
Together with Part 1.\ we simply have $p_+=p_-$, and $K^E_\cC(p_\pm)=K(p_\pm)$.
\item
The radial momentum components do not appear in $p_\pm$ 
but can be reconstructed from the energy condition, since their energy
equals $E_\pm - K^E_\cC(p_\pm)$.
As $\|q_\pm^I \|_\cM = 1$, the inverse square roots in \eqref{def:L:CEE} are the 
times of collision with $\Delta_\cC$.
\hfill $\Diamond$
\end{enumerate}

\end{remark}
\textbf{Proof of  Lemma \ref{lem:Lag:rel}:}
As $P_\cC = T^*Z_\cC \cong T^*\bS^I_\cC \times T^* \Delta^E_\cC$, we note that
\begin{enumerate}[$\bullet$]
\item 
by the conditions  $(p_-)^I_\cC=(p_+)^I_\cC=0$ on $T^*\bS^I_\cC\times T^*\bS^I_\cC$, 
we consider the zero section of the bundle 
$T^*(\bS^I_\cC\times \bS^I_\cC)\to \bS^I_\cC\times \bS^I_\cC$. But zero sections
$\iota: N\to T^*N$ are Lagrangian, since for the tautological one-form $\theta$ on $T^*N$
one has $\iota^*\theta=0$. So for the symplectic two-form $\omega=-d\theta$ we have
$\iota^*\omega =-\iota^*d\theta=-d\iota^* \theta=0$;
\item 
the linear independent conditions  $(p_+)^E_\cC = (p_-)^E_\cC$ and, with the internal
kinetic energies $E_\pm - K^E_\cC(p_\pm)>0$, 
equality of collision points on $\Delta^E_\cC$:
\beq
{\textstyle (q_+)^E_\cC - 
\frac{(\cM^{-1} p_+)^E_\cC}{\sqrt{2(E_+ - K^E_\cC(p_+))}} = 
(q_-)^E_\cC + \frac{(\cM^{-1} p_-)^E_\cC}{\sqrt{2(E_- - K^E_\cC(p_-))}}}
\Leq{equal:external:Delta}
define a Lagrangian subspace of $T^* \Delta^E_\cC\times T^* \Delta^E_\cC$. 
Note therefore that the coordinate change on a suitable subset of 
$T^* \Delta^E_\cC\times T^* \Delta^E_\cC$
\begin{align*}
& \big( (p_+)^E_\cC\,,\, (q_+)^E_\cC \,,\, (p_-)^E_\cC\,,\, (q_-)^E_\cC \big) \longmapsto \\ 
&\textstyle
\Big( (p_+)^E_\cC \,,\, (q_+)^E_\cC - 
\frac{(\cM^{-1} p_+)^E_\cC}{\sqrt{2(E_+ - K^E_\cC(p_+))}} \,,\, (p_-)^E_\cC \,,\,
(q_-)^E_\cC + \frac{(\cM^{-1} p_-)^E_\cC}{\sqrt{2(E_- - K^E_\cC(p_-))}} \Big)
\end{align*}
is a symplectomorphism and in these new coordinates it is apparent that the conditions 
define a Lagrangian subspace.
\hfill $\Box$\\
\end{enumerate}
Note that $L_\cC$ is not the graph of a map $P_\cC\to P_\cC$.
We want to compose such Lagrangian relations $L_{\cC;E_1,E_2}$ and 
$L_{\cD;E_2,E_3}$ (with an additional connecting Lagrangian relation
\eqref{Lagrangian:additional}).

\subsection{Scattering by several subspaces}
%

We now want to compose scattering by different subspaces. This amounts to
composition of Lagrangian relations. To set the stage, 
we first consider compositions of relations in different categories.
This has been discussed by {\sc Weinstein} in \cite{We1,We2}.
\begin{remark}[composition of relations]\quad\\[-6mm] 
\begin{enumerate}[1.]
\item 
For {\bf sets} $X,Y,Z$ the {\em composition} of relations $L\subseteq X\times Y$ and 
$M\subseteq Y\times Z$ is given by the relation 
(which we write as for the case of composition of maps)
\beq 
M\circ L := \{ (x,z)\in X\times Z\mid \exists\, y\in Y: (x,y)\in L\mbox{ and }(y,z)\in M\} \, .
\Leq{def:composition}
\item 
For {\bf $\mathbb K$-vector spaces} a relation is called {\em linear} if it is a 
linear subspace of the respective 
product space. The composition of linear relations is a linear relation. Even if
$X$, $Y$, $Z$, $L$ and $M$ are of the same finite dimension 
(as it is the case for the Lagrangian relations we consider), 
$\dim(M\circ L)$ can be different if $L$ or $M$ is not a graph. 
As examples, consider 
$L:=X\times\{0\}$ and $M:=\{0\}\times Z$ with $M\circ L =X\times Z$, or
$L:=\{0\}\times Y$ and $M:=Y\times\{0\}$ with $M\circ L=\{0\}\times \{0\}$.
This defect is healed by assuming strong transversality.
\item 
However, as shown in \cite[Section 4]{GS} of {\sc Guillemin} and {\sc Sternberg}, 
the composition of 
{\bf linear Lagrangian relations} is a linear Lagrangian relation. 
\item 
{\bf Smooth relations} $L,M$, {\em i.e.}\  submanifolds, are called {\em transverse} if
$L\times M$ is transverse to $X\times \Delta_Y\times Z$, with $\Delta_Y:=\{(y,y)\mid y\in Y\}$.
Then their intersection $L\times_Y\! M := 
(L\times M)\cap(X\times \Delta_Y\times Z)$ 
can be considered as a submanifold of $X\times Y\times Z$. 
$L,M$ are called
{\em strongly transverse} if additionally the projection $L\times_Y\! M \to X\times Z$,
$(x,y,z)\mapsto (x,z)$ is an embedding of $M\circ L$. 
We denote this by $M\pitchfork L$.
For a closed embedding this is true iff
\begin{enumerate}[a)]
\item 
$(M,L)$ is {\em monic},  {\em i.e.} for $(x,z)\in M\circ L$ there exists a unique $y$ in 
\eqref{def:composition},
\item 
$(TM,TL)$ is {\em monic},
\item 
The map $M\times_Y L\to M\circ L$ is proper,
\end{enumerate}
since\, a) and b)\, is equivalent to that map being an injective immersion, and since a proper 
map with a locally compact Hausdorff target space is closed.
In that case $M\circ L$ is a smooth relation,
see \cite{We1} and \cite[Definition 3.3]{We2}.
\item 
For general {\bf (smooth) Lagrangian relations} $L\subseteq X\times Y$ and 
$M\subseteq Y\times Z$ {\em strong transversality} ensures that 
$M\circ L$ is a smooth Lagrangian relation.
\hfill $\Diamond$
\end{enumerate}
\end{remark}

We now use the above theory to consider more than one collision.\\
In Lemma \ref{lem:composition} we first prove directly strong transversality for the 
composition of only {\em two} Lagrangian relations.
As this already turns out to be involved, when we compose {\em several} such relations,
see \eqref{finite:seq}, in Theorem \ref{thm:severel:relations} we switch to an approach 
based on the minimisation of a functional.\footnote{Although we redo the case
of two relations in Lemma \ref{lem:N:2:3}, using our functional, we keep 
Lemma \ref{lem:composition},
so that one can better compare the two methods.}\\ 

\begin{figure}[h]
\centerline{\includegraphics[width=70mm]{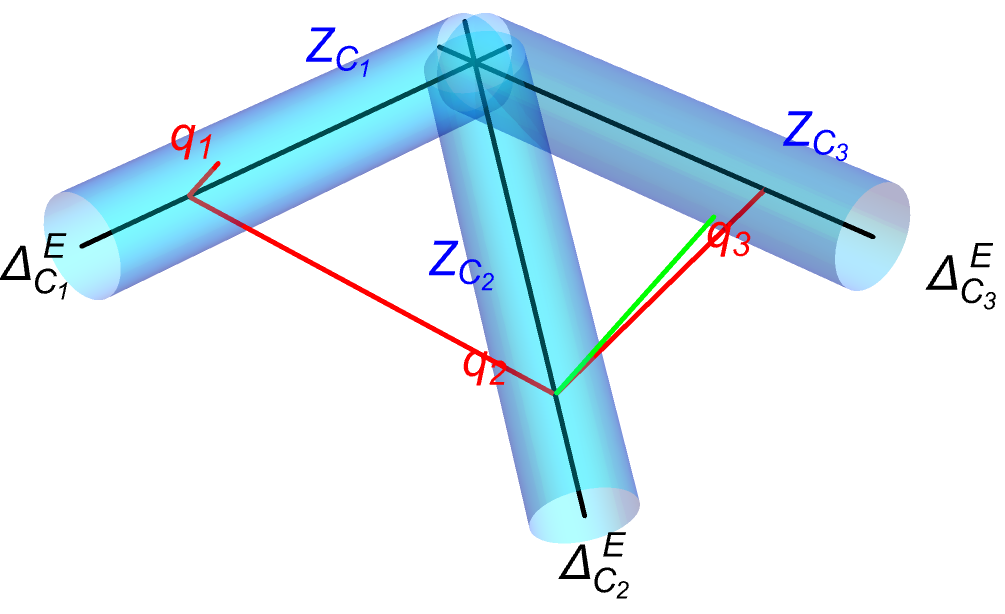}}
\caption{Solutions in the composition 
$L_{\cC_2,\cC_3;E_2,E_3} \circ L_{\cC_1,\cC_2;E_1,E_2}$ of
Lagrangian relations that intersect 
$\Delta_{\cC_3}$ (red), respectively do not (green)}
\label{fig:lagrangianIntersection}
\end{figure}

For partitions $\cC\neq \cD\in \cP_\Delta (N)$ consider the 
three submanifolds $Y_\cC^-, Y_\cC^+$ and $Y_\cD^-$. 
We have defined in Lemma~\ref{lem:Lag:rel} a Lagrangian relation in
$\cP_\cC \times \cP_\cC $, with $\cP_\cC  \cong Y_\cC^\pm / \! \sim$. 
Now some points in $Y_\cC^+$ will hit $Y_\cD^-$ 
when evolving according to the free flow. 
We denote the (open) subset of these points by $\widetilde{Y}_{\cC,\cD}^+$, and its 
image by $\widetilde{Y}_{\cC,\cD}^-\subseteq Y_\cD^-$.
The free flow $\Phi$, see \eqref{free:flow}, defines Poincar\'e maps, 
more precisely diffeomorphisms
\beq
\mbox{\rm Poi}_{\cC,\cD}: \widetilde{Y}_{\cC,\cD}^+ \to \widetilde{Y}_{\cC,\cD}^- \qmbox{,}
x\mapsto \Phi\big(T_{\cC,\cD}(x),x\big) 
\qquad
\big(\cC\neq \cD\in \cP_\Delta (N)\big) 
\Leq{def:poinc} 
with the relatively open submanifolds $\widetilde{Y}_{\cC,\cD}^+\subseteq Y_\cC^+$
and $\widetilde{Y}_{\cC,\cD}^-\subseteq Y_\cD^-$ 
such that the Poincar\'e time $T_{\cC,\cD}$ is positive. 
By strict convexity of 
\[t\mapsto \| q^I_\cE(t)\|_\cM^2= \| q^I_\cE(0)+\cM^{-1}p^I_\cE(0)t\|_\cM^2 \qquad 
\big( \cE\in \cP_\Delta(N) \big)\] 
and transversality of the flow to the Poincar\'e surfaces $ \widetilde{Y}_{\cC,\cD}^\pm$,
the Poincar\'e times $T_{\cC,\cD}$ are uniquely defined and smooth.\\
As $\Phi$ preserves the kinetic energy $K$, the $\mbox{\rm Poi}_{\cC,\cD}$ 
descend for every $E>0$ to Lagrangian relations
\beq 
L_{\cC,\cD;E} \subseteq  \cP_\cC\times \cP_\cD \, ,
\Leq{Lagrangian:additional}
so that $\dim(L_{\cC,\cD;E}) = 2(D-1)$.
By Lemma \ref{lem:Lag:rel} the composition
\beq
L_{\cC,\cD;E_-,E_+} 
:= L_{\cC,\cD;E_+} \circ L_{\cC;E_-,E_+}\subseteq  \cP_\cC\times \cP_\cD
\Leq{L:CDEE}
is trivially a Lagrangian relation,
since $L_{\cC,\cD;E_+}$ is the graph of a symplectomorphism 
(see \cite[Section 2]{We1}).
However, $L_{\cC,\cD;E_-,E_+}$ is not the graph of a map.
We can now compose Lagrangian relations of the type \eqref{L:CDEE},
see Figure \ref{fig:lagrangianIntersection}.

In order to obtain a Lagrangian relation, we have to check strong transversality.
In Lemma \ref{lem:composition} we first do this for a {\em single} such composition. 
We find its proof instructive, although, strictly speaking, it is not needed, since 
afterwards we will use a minimisation method
to show strong transversality for {\em finitely many} compositions.
\begin{lemma}[composition of two Lagrangian relations] \quad
\label{lem:composition}\\
The strong transversality conditions
\[ L_{\cC_2,\cC_3;E_2,E_3} \pitchfork L_{\cC_1,\cC_2;E_1,E_2}
\qquad\big(\cC_1\neq \cC_2\neq \cC_3\in \cP_\Delta (N),\ E_1,E_2,E_3>0\big)\]
hold true. So $L_{\cC_2,\cC_3;E_2,E_3} \circ L_{\cC_1,\cC_2;E_1,E_2} \subseteq
\cP_{\cC_1}\times \cP_{\cC_3}$ 
are Lagrangian relations.
\end{lemma}
\textbf{Proof:}\\[-6mm]
\begin{enumerate}[$\bullet$]
\item 
{\bf The pair $(L_{\cC_2,\cC_3;E_2,E_3}, L_{\cC_1,\cC_2;E_1,E_2})$ is transverse:}\\
We must prove that for all 
$x_k=(p_k,q_k)\in \cP_{\cC_k}$ with 
$(x_1,x_2)\in L_{\cC_1,\cC_2;E_1,E_2}$
and $(x_2,x_3)\in L_{\cC_2,\cC_3;E_2,E_3}$ 
(so that $(x_1,x_3) \in L_{\cC_2,\cC_3;E_2,E_3} \circ L_{\cC_1,\cC_2;E_1,E_2}$)
we have
\beq
T_{x_2} P_{\cC_2} = 
 T\pi_1(T_{(x_2,x_3)} L_{\cC_2,\cC_3;E_2,E_3}) +
 T\pi_2 \big(T_{(x_1,x_2)} L_{\cC_1,\cC_2;E_1,E_2} \big)\, ,
\Leq{transversality:cond}
with the projection $\pi_k$ on the $k$--th factor in \eqref{L:CDEE}.
We can split the left hand side in the internal and external part:
$T_{x_2} P_{\cC_2} = T_{(0, q^I_2 )} T^*\bS^I_{\cC_2} \oplus  
T_{(p^E_2, q^E_2 )} T^*\Delta^E_{\cC_2}$. 

Furthermore, we choose the following linear subspaces:
\begin{enumerate}[$-$]
\item 
Concerning $(x_1,x_2)\in L_{\cC_1,\cC_2;E_1,E_2}$, with $x_k = (p_k,q_k)$,
\[ \hspace*{-5mm}
S_1:= \big\{(\delta x_1,\delta x_2) \in T_{(x_1,x_2)} L_{\cC_1,\cC_2;E_1,E_2} \mid 
\textstyle
\delta \big( q_1^E + \frac{(\cM^{-1} p_1)^E}{\sqrt{2(E_1 - K^E_{\cC_1}(p_1^E))}} \big)=0
, \delta q_1^I =0
\big\}  \]
(compare with Definition \eqref{def:L:CEE})
defines a family of rays focussing on 
the point $Q_1\in \Delta_{\cC_1}^E$ implicitly given by $(x_1,x_2)$ 
and defocussing afterwards, compare Figure~\ref{fig:FocussingDefocussingRays}.
\begin{figure}[htbp]
\centerline{\includegraphics[width=70mm]{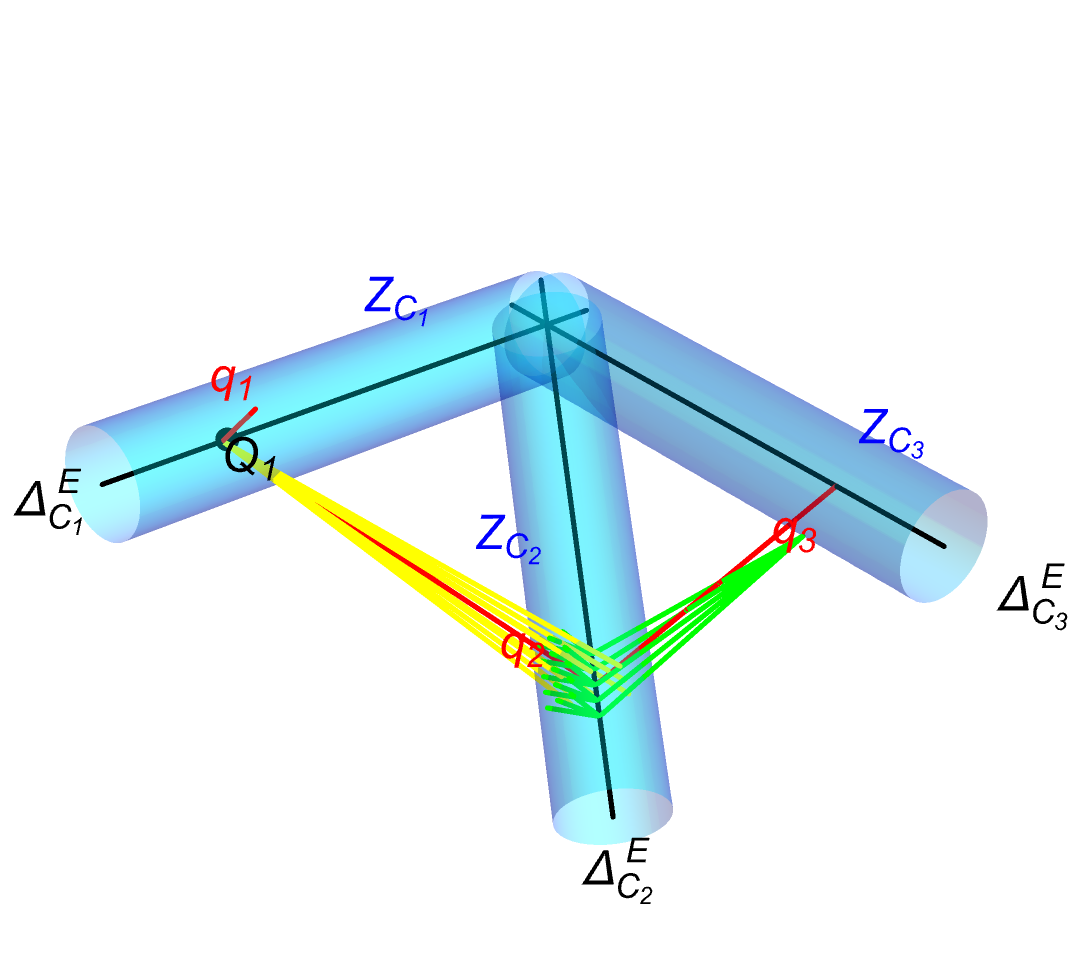}}
\caption{The rays (yellow) with focus $Q_1$ and the rays (green) with focus $q_3$.}
\label{fig:FocussingDefocussingRays}
\end{figure}

\item 
Concerning $(x_2,x_3) \in L_{\cC_2,\cC_3;E_2,E_3}$, 
\[ S_3:= \{(\delta x_2,\delta x_3) \in T_{(x_2,x_3)} L_{\cC_2,\cC_3;E_2,E_3} \mid 
\delta q_3=0 \}  \]  
defines a family of rays focussing on $q_3$.
\end{enumerate}
These families of defocussing respectively focussing rays are both of dimension 
$D-1 = \eh \dim(T_{x_2} P_{\cC_2})$, 
since Poincar\'e times were assumed to be positive
in \eqref{def:poinc}.  
Our claim that (with the projections $\pi_k$ to the $k$--th factor of 
$L_{\cC_\ell,\cC_{\ell+1};E_\ell,{E_{\ell+1}}} \subseteq  
\cP_{\cC_\ell} \times \cP_{\cC_{\ell+1}}$)
\beq
T_{x_2} P_{\cC_2} = T\pi_1(S_3) \oplus T\pi_2(S_1)
\Leq{just:transversal}
of course implies the transversality condition \eqref{transversality:cond}.

Since $T_{q_2}Z_{\cC_2}$ is not necessarily perpendicular to the velocity vector 
$\cM^{-1}p_2$, we consider the auxiliary linear subspace 
\[ X := \{\delta q \in T_{q_2}M \mid \langle p_2, \delta q \rangle = 0 \} \]
(which is also of dimension $D-1$). 
Then, with the time $t > 0$ needed for the trajectory to go from 
$Q_1 \in \Delta_{\cC_1}$ to $q_2\in Z_{\cC_2}$, 
the family of defocussing rays forms the linear Lagrangian subspace 
\[U^+ := \big\{ \big(t^{-1}\cM\delta q,  \delta q \big) \in T^*X \mid \delta q\in X\big\} \] 
of dimension $D-1$. So the restriction of the quadratic form 
\[ T^*X \to \bR\quad,\quad 
(\delta p,\delta q) \mapsto \langle \delta p,\delta q\rangle\] 
to $U^+$ is positive definite. Similarly, we obtain from the family of focussing rays a 
$(D-1)$-dimensional subspace $U^-$ of  $T^*X$ on which that quadratic form 
is negative definite. A dimension count shows that $T^*X = U^+ \oplus U^-$.\\
However, the linear symplectomorphism $T^*X\to T_{x_2} P_{\cC_2}$ induced by 
the projection $X\to T_{q_2}Z_{\cC_2}$ along 
the velocity vector $\cM^{-1}p_2^-$ maps $U^+$ to
$T\pi_2(S_1)$ and $U^-$ to $T\pi_1(S_3)$. This proves \eqref{just:transversal}. 
\item 
{\bf The pair $(L_{\cC_2,\cC_3;E_2,E_3}, L_{\cC_1,\cC_2;E_1,E_2})$ is monic:}\\
Let $x_k=(p_k,q_k)\in \cP_{\cC_k}$ be such that  $(x_1,x_2)\in L_{\cC_1,\cC_2;E_1,E_2}$
and $(x_2,x_3)\in L_{\cC_2,\cC_3;E_2,E_3}$.
For $\tilde{x}_2 \equiv (\tilde{p}_2,\tilde{q}_2)\in \cP_{\cC_2}$ with 
$(x_1,\tilde{x}_2)\in L_{\cC_1,\cC_2;E_1,E_2}$ and 
$(\tilde{x}_2,x_3)\in L_{\cC_2,\cC_3;E_2,E_3}$ we have $\tilde{x}_2=x_2$ since,
using Definition \eqref{def:L:CEE} and conservation of momentum for 
the Poincar\'e maps $\mbox{\rm Poi}_{\cC,\cD}$
\begin{enumerate}[1.]
\item 
By conservation of momentum, $\tilde{p}_{2,+} =  p_{3,-} \equiv p_3 = p_{2,+}$; so 
\begin{enumerate}[--]
\item 
external momenta coincide:
\[(\tilde{p}_2)^E_{\cC_2} \!\equiv \! (\tilde{p}_{2,-})^E_{\cC_2}  
\! = \! (\tilde{p}_{2,+})^E_{\cC_2} \! =  \!(p_{2,+})^E_{\cC_2} \! =  \! (p_2)^E_{\cC_2} \, . \]
\item 
Outgoing internal momenta coincide: $(\tilde{p}_{2,+})^I_{\cC_2} = (p_{2,+})^I_{\cC_2}$.
\end{enumerate}
\item 
So the collision points $k_{\cC_2}$ with $\Delta^E_{\cC_2}$ coincide, 
since both trajectories have the point $q_3$ in common.
\item 
From $x_1$ we can reconstruct the collision point $k_{\cC_1}$ with $\Delta^E_{\cC_1}$ 
that both trajectories have in common. 
But then we get the same path between $k_{\cC_1}$ and $k_{\cC_2}$ and from this we 
can reconstruct the points $q_2$ (and $\tilde{q}_2$) as the intersection of this path with 
$\cP_{\cC_2}$. 
Additionally we obtain the velocities of all particles as the slope of the line, 
from which we can determine $p_2$.
\end{enumerate}
\item 
{\bf The pair $(TL_{\cC_2,\cC_3;E_2,E_3}, TL_{\cC_1,\cC_2;E_1,E_2})$ is monic:}\\
Consider for $k=1,2,3$ some points $(x_k, y_k) \in T P_{\cC_k}$ with $x_k = (p_k, q_k)$, 
$y_k = (\delta p_k, \delta q_k)$ such that for $k=1,2$ we have 
$((x_k, y_k), (x_{k+1}, y_{k+1})) \in TL_{\cC_k, \cC_{k+1}; E_k, E_{k+1}}$. 
We will prove now that we can reconstruct $(x_2, y_2)$ from $(x_1, y_1)$ and $(x_3, y_3)$, 
which proves the claim. \\
First of all, $x_2$ is uniquely determined from $x_1$ and $x_3$ by the previous step of the 
proof. Hence, we only have to care about the tangent vector $y_2$. 
Therefor, we consider two families of curves 
$\eps \mapsto (\tilde{x}_1^{(\eps)}, \tilde{x}_2^{(\eps)}) \in L_{\cC_1, \cC_{2}; E_1, E_{2}}$ 
and $\eps \mapsto (\hat{x}_2^{(\eps)}, \tilde{x}_3^{(\eps)}) \in L_{\cC_2, \cC_{3}; E_2, E_{3}}$ 
such that for $k=1,2,3$ we have 
$\tilde{x}_k^{(\eps)} = x_k + \eps y_k + o(\eps)$ and $\hat{x}_2^{(\eps)} = x_2 + \eps y_2 + 
o(\eps)$. 
Note that we can have two different curves 
$\tilde{x}_2^{(\eps)}$, $\hat{x}_2^{(\eps)}$ on $T P_{\cC_2}$. 

We will construct now a point $x_2^{(\eps)}$ just with 
$\tilde{x}_1^{(\eps)}$ and $\tilde{x}_3^{(\eps)}$ that also satisfies 
$x_2^{(\eps)} = x_2 + \eps y_2 + o(\eps)$, hence we can reconstruct $y_2$ from $y_1$ 
and $y_3$. Therefor, consider the point $\tilde{Q}_1^{(\eps)} \in \Delta_{\cC_1}$ on the
corresponding trajectory starting at $\tilde{q}_1^{(\eps)}$ with direction determined by 
$\tilde{p}_1^{(\eps)}$, compare Figure~\ref{fig:FocussingDefocussingRays}. 
The trajectory has to intersect $\Delta_{\cC_1}$, as by assumption 
$(\tilde{x}_1^{(\eps)}, \tilde{x}_2^{(\eps)}) \in L_{\cC_1, \cC_{2}; E_1, E_{2}}$. 
Analogously there is a point $\tilde{Q}_2^{(\eps)} \in \Delta_{\cC_2}$ on the corresponding 
trajectory of $(\hat{x}_2^{(\eps)}, \tilde{x}_3^{(\eps)})$, but we can again determine this point 
solely by $\tilde{x}_3^{(\eps)}$. 
Now we can consider the line segment through $\tilde{Q}_1^{(\eps)}$ and 
$\tilde{Q}_2^{(\eps)}$ and its intersection with $Z_{\cC_2}$, which we call $q_2^{(\eps)}$. 
From the slope of the line we can also determine $p_2^{(\eps)}$. In order to prove that for this 
choice we indeed have $x_2^{(\eps)} = x_2 + \eps y_2 + o(\eps)$, we compare the line 
segment with the line through $\tilde{Q}_1^{(\eps)}$ and $\tilde{q}_2^{(\eps)}$ and the line 
through $\tilde{Q}_2^{(\eps)}$ and $\hat{q}_2^{(\eps)}$:
\begin{itemize}
\item 
The slope of these auxiliary lines are determined by $\tilde{p}_2^{(\eps)}$ resp. 
$\hat{p}_2^{(\eps)}$, hence they differ only in terms of size $o(\eps)$.
\item 
The lines are close at least in the size of $o(\eps)$, as this is the distance between 
$\tilde{q}_2^{(\eps)}$ resp. $\hat{q}_2^{(\eps)}$.
\end{itemize}
From this one can deduce that the same holds true for the line through 
$\tilde{Q}_1^{(\eps)}$ and $\tilde{Q}_2^{(\eps)}$, so the claim follows.
\item 
{\bf $\Pi: L_{\cC_2,\cC_3;E_2,E_3}\times_{\cP_{\cC_2}} L_{\cC_1,\cC_2;E_1,E_2} \to 
L_{\cC_2,\cC_3;E_2,E_3}\circ L_{\cC_1,\cC_2;E_1,E_2}$ is proper:}\\[1mm]
Choose a compact
$K \subseteq L_{\cC_2,\cC_3;E_2,E_3}\circ L_{\cC_1,\cC_2;E_1,E_2}$.
\begin{enumerate}[1.]
\item 
Then its preimage $\Pi^{-1}(K)$ is closed, since $\Pi$ is continuous.
\item 
For data on $K$ the positions $q_1$ and $q_3$ are bounded.
But $\| q_{2} \|_\cM \le \max(\| q_{1} \|_\cM ,\| q_{3} \|_\cM )$
by convexity of $J$ (Theorem \ref{thm:finer}.2).

The momentum $p_2$ is bounded, too, since kinetic energy $K(p_2)$ is bounded
above by $E_2$. 
\end{enumerate}
As $\Pi^{-1}(K)$ is closed and bounded, it is compact.
\hfill $\Box$
\end{enumerate}
When we compose (as sets) a finite sequence of smooth Lagrangian relations 
\beq
L_{\cC_k,\cC_{k+1};E_k,E_{k+1}}\qquad (k=1,\ldots,N-1)
\Leq{finite:seq}
with $\cC_{k+1} \neq \cC_k$, we shall prove that the resulting relation is again 
smoothly Lagrangian, after removing finitely many sets of codimension $d$.
For this, it is useful to relate the composition to the set of critical points
of the functionals 
\beq
I_{q_N,q_1}: \Delta^E_{\cC_{N-1}}\times \ldots \times \Delta^E_{\cC_2} \to \bR\
\mbox{,} \quad  
(q_{N-1},\ldots,q_2) 
\mapsto \sum_{k=1}^{N-1} \sqrt{E_{k+1}}\, \|q_{k+1} - q_k\|_\cM  \, . 
\Leq{I:functional}
$I_{q_N,q_1}$ is continuous and convex (though not strictly convex, 
see Remark \ref{rem:nonuniqueness}).
It is smooth at points with $q_{k+1} \neq q_k$.  
By the coercivity property $\lim_{\|q\|\to \infty} I_{q_N,q_1}(q) = \infty$,
a minimiser $q$ always exists. The problems are its uniqueness and, if so, its
dependence on $q_1$ and $q_N$. 
We start by considering the simplest cases.
\begin{lemma}[cases $N=2$ and $N=3$] \quad\\[-6mm] \label{lem:N:2:3}
\begin{enumerate}[1.]
\item 
For $I: \Delta^E_{\cC_2}\times \Delta^E_{\cC_1}\to \bR$, 
$I(q_2,q_{1}) = \|q_1-q_2\|_\cM$, $q_1 \neq q_2$ and 
$\hat{q}_{1,2} := \frac{q_1-q_2}{\|q_1-q_2\|_\cM}$
\[D_{(q_2,q_1)} I(\delta q_{2},\delta q_{1})  = 
\langle \hat{q}_{1,2},\delta q_{1} - \delta q_{2} \rangle_\cM \ \quad
\big( (\delta q_{2},\delta q_{1})
\in T_{q_2} \Delta^E_{\cC_2} \oplus T_{q_1} \Delta^E_{\cC_1} \big)\]
and $D^2_{(q_2,q_1)} I $ is the positive semidefinite bilinear form on 
$T_{q_2} \Delta^E_{\cC_2} \oplus T_{q_1} \Delta^E_{\cC_1}$
\[D^2_{(q_2,q_1)} I 
\big(\delta q^{I}_{2},\delta q^{I}_{1}\,;\,\delta q^{I\!I}_{2},\delta q^{I\!I}_{1} \big)
= \frac{ \big\langle \delta q^{I}_{1}-\delta q^{I}_{2} \,, \,(\idty-P_{\hat{q}_{1,2}}) \,
(\delta q^{I\!I}_{1}-\delta q^{I\!I}_{2})  \big\rangle_{\!\cM}}{ \|q_1 - q_2\|_\cM} \, ,\]
with $P_{\hat{q}}$ denoting the orthogonal projection to $\mathrm{span}(\hat{q})$.
\item 
For $q_k\in \Delta^E_{\cC_k}\backslash \Delta^E_{\cC_2}$ $(k=1,3)$ the minimiser $q_2$
of $I_{q_3,q_1}:  \Delta^E_{\cC_2}\to \bR$ exists uniquely and is a critical point.
It corresponds to the condition that the ingoing and
outgoing $\cC_2$--external momenta coincide.
\item 
The minimiser 
$Q: (\Delta^E_{\cC_3}\backslash  \Delta^E_{\cC_2})\times 
(\Delta^E_{\cC_1} \backslash \Delta^E_{\cC_2}) \to \Delta^E_{\cC_2}$ 
in Part 2.\ is smooth.
\end{enumerate}
\end{lemma}
\textbf{Proof:}
\begin{enumerate}[1.]
\item 
is a concrete calculation.
\item 
By Item 1., for $q_2\in \Xi_{\cC_2}$
\[\textstyle
D_{q_2} I_{q_3,q_1}(\delta q_2) = 
\langle \sqrt{E_2}\, \hat{q}_{2,3}- \sqrt{E_1}\,\hat{q}_{1,2} ,\delta q_{2} \rangle_\cM 
\qquad \big( \delta q_{2} \in T_{q_2} \Delta^E_{\cC_2} \big) .\]
As momenta scale with the square root of the kinetic energies $E_k$, 
$D_{q_2} I_{q_3,q_1}=0$ iff 
ingoing and outgoing $\cC_2$--external components of the momenta $p_\pm$ with  
directions $\cM^{-1}p_- /\|p_-\|_{\cM^{-1}} = \hat{q}_{1,2}$ and 
$\cM^{-1}p_+/\|p_+\|_{\cM^{-1}} = \hat{q}_{2,3}$ coincide.

The bilinear form equals $D^2_{q_2} I_{q_3,q_1}(\delta q^{I}_2,\delta q^{I\!I}_2) =$
\beq 
\frac{\sqrt{E_1}\, \big\langle \delta q^{I}_{2} \,, \,(\idty-P_{\hat{q}_{1,2}}) \,
\delta q^{I\!I}_{2} \big\rangle_\cM}{ \|q_1 - q_2\|_\cM } \,+\,
\frac{\sqrt{E_2}\, \big\langle \delta q^{I}_{2} \,, \,(\idty-P_{\hat{q}_{2,3}}) \,
\delta q^{I\!I}_{2} \big\rangle_\cM}{ \|q_2 - q_3\|_\cM }\,.
\Leq{D:2:q2}
So if $\hat{q}_{1,2} \not\in \Delta^E_{\cC_2}$, or equivalently if
$q_1\not\in \Delta^E_{\cC_2}$ 
(which follows from our condition 
$q_1\in  \Delta^E_{\cC_1} \backslash \Delta^E_{\cC_2}$), 
then this bilinear form on $T_{q_2} \Delta^E_{\cC_2}$ 
is positive definite, implying uniqueness of the critical point. 
The same is true for $q_3\in  \Delta^E_{\cC_3} \backslash \Delta^E_{\cC_2}$. 

In the c.o.m.\ subspace $\Delta^E_{\cC_{\max}} = \{0\}$ so that for  
$\cC_2 = \cC_{\max}$ the critical point
is trivially unique. 
As $\dim(\Delta^E_\cC)=d(|\cC|-1)$, see \eqref{dim:MC}, 
the dimension of $\Delta^E_{\cC_2}$ is positive otherwise. 
As $\lim_{\Delta^E_{\cC_2} \ni\, q_2\to \infty} I_{q_3,q_1}(q_2) = \infty$ 
and $I_{q_3,q_1}$ is strictly convex, there exists a unique minimiser 
$q_2\in \Delta^E_{\cC_2}$ (but not necessarily in $\Xi^{(0)}_{\cC_2}$). 
\item
Smoothness of $Q$ follows from the implicit function theorem, applied to the 
map $F:(\Delta^E_{\cC_3}\times\Delta^E_{\cC_1}) \times \Delta^E_{\cC_2} \to 
\Delta^E_{\cC_2}$,
$F(q_3,q_1;q_2):= D_{q_2} I_{q_3,q_1}$, 
since the minimiser of  $I_{q_3,q_1}$
is given by the unique solution of $q_2\mapsto F(q_3,q_1;q_2)=0$,
using regularity of \eqref{D:2:q2}. Concretely,
\[DQ(q_3,q_1) = - (D^2_{Q(q_3,q_1)} I_{q_3,q_1})^{-1} D_{q_3,q_1}D_{Q(q_3,q_1)} 
I_{q_3,q_1} \, . \hfill \tag*{$\Box$}\]
\end{enumerate}
\begin{remark}[non-uniqueness of minimisers] \label{rem:nonuniqueness}\quad\\
For $E_1=E_2$ and (contrary to the assumption of Lemma \ref{lem:N:2:3}.2) 
$q_1,q_3\in \Delta^E_{\cC_2}$,
all convex combinations
$q_2$ of $q_1$ and $q_3$ are trivially minimisers of $I_{q_3,q_1}$. So for
$q_1\neq q_3$ the minimiser would not be unique. \\
By a variation of that argument one sees that then the minimiser $Q$ in Part 3 of
Lemma \ref{lem:N:2:3} would not be Lipschitz continuous.
\hfill $\Diamond$
\end{remark}

For $N>3$ the functional $I_{q_N,q_1}$ is not differentiable on the full domain
$\Delta^E_{\cC_{N-1}} \times \ldots \times \Delta^E_{\cC_{2}}$. 
Hence (unlike for the case $N=3$ of Lemma \ref{lem:N:2:3}) unique existence
of a minimiser cannot be expected. By restricting the domain, we still can show uniqueness:
\begin{theorem} \quad\\ \label{thm:severel:relations}
For $N\in \bN$, arbitrary $q_1\in \Xi_{\cC_1}$, $q_N\in \Xi_{\cC_N}$, 
minimisers of $I_{q_N,q_1}$ on the open dense submanifold
$\Xi_{\cC_{N-1}} \times \ldots \times \Xi_{\cC_2}$ are unique for 
$\cC_1,\ldots,\cC_N\in \cP_\Delta$ with $\cC_{k+1}\neq \cC_k$.
\end{theorem}
\textbf{Proof:}
For the sequence \eqref{finite:seq} of $\cC_k$ and
$q_k \in \Xi_{\cC_k}$ $(k=1,\ldots,N)$, see \eqref{Xi:Null},  we have $q_{k+1}\neq q_k$.
Then $I_{q_N,q_1}$  is smooth at $q=(q_{N-1},\ldots,q_{2})$, with
\begin{align*}
D_q  I_{q_N,q_1} (\delta q) =& \,  
\sqrt{E_{N-1}} \, \langle \hat{q}_{N-1,N} , \delta q_{N-1} \rangle_\cM
- \sqrt{E_1}\, \langle \hat{q}_{1,2} , \delta q_2 \rangle_\cM \\ 
&\textstyle + \sum_{\ell=2}^{N-2} 
\sqrt{E_\ell} \, \langle \hat{q}_{\ell,\ell+1} , \delta q_{\ell+1} - \delta q_\ell \rangle_\cM  
\end{align*}
and
\begin{align*}
D^2_q  I_{q_N,q_1} (\delta q^I, \delta q^{I\!I}) =& \textstyle\,
\sqrt{E_1}\, \langle \delta q^I_2,
\frac{\idty-P_{\hat{q}_{1,2}}}{\| q_1 - q_2 \|_\cM} \delta q^{I\!I}_2 \rangle_\cM \\
&+ \textstyle \sqrt{E_{N-1}} \, \langle \delta q^I_{N-1},
\frac{\idty-P_{\hat{q}_{N-1,N}}}{\| q_{N-1} - q_N \|_\cM} \delta q^{I\!I}_{N-1} \rangle_\cM\\
&+\textstyle \sum_{\ell=2}^{N-2}  \sqrt{E_\ell} \, 
\langle \delta q^I_\ell  - \delta q^I_{\ell+1} , 
\frac{\idty - P_{\hat{q}_{\ell,\ell+1}}}{\| q_\ell - q_{\ell+1} \|_\cM} 
(\delta q^{I\!I}_\ell  - \delta q^{I\!I}_{\ell+1}) \rangle_\cM \, .
\end{align*}
All terms are obviously positive semidefinite. 
In fact $D^2_q  I_{q_N,q_1}$ is positive definite, since 
for $\delta q^I = \delta q^{I\!I}$ the third term only vanishes if 
$\delta q^I_2 = \ldots = \delta q^I_{N-1}$, and then the first two terms only vanish if 
$\delta q^I_2=0$.\\
Since $I_{q_N,q_1}: \Delta^E_{\cC_{N-1}} \times \ldots \times \Delta^E_{\cC_{2}} \to \bR$ 
is convex, this property suffices to show uniqueness of a minimiser.
\bigskip
For arbitrary $N\ge3$,
$D^2_{q}I_{q_N,q_1} \ge 0$ has a matrix representation of the form 
\[\bsm 
M_1+M_2 &-M_2&0&0&\ldots&0\\
-M_2& M_2+M_3&-M_3&0&\ldots&0\\
0     & -M_3& M_3+M_4&-M_4&\ldots&0\\
\vdots&&&&\vdots\\ 
0&\ldots&&0& -M_{N-2}&M_{N-2}+M_{N-1}
\esm\]
with $M_k:= \sqrt{E_k}(\idty-P_{\hat{q}_{k,k+1}})/\|q_{k+1}-q_k\|_\cM$. 
So 
\begin{enumerate}[$\bullet$]
\item 
this symmetric matrix has positive diagonal entries. 
\item 
It is {\em weakly diagonally dominant}, meaning that in each line the sum of the off-diagonals 
is smaller or equal in absolute value than the diagonal.
\item 
Furthermore, there exist lines (the first and the last one), where this inequality is strict.
\item
It is also {\em irreducible}, meaning that it cannot be conjugated by permutation matrices
to a matrix of the block form $\bsm L_{1,1}&0\\ L_{2,1}&L_{2,2}\esm$.
This can be inferred from the directed graph of the matrix
(whose adjacency matrix is the indicator of the non-zero entries of the matrix), 
which is strongly connected.
\end{enumerate}
These properties imply that the matrix is positive definite, see {\sc Horn} and 
{\sc Johnson} \cite[Cor. 6.2.9]{HJ}.
\hfill $\Box$

\begin{remark}[Comparison with Lemma \ref{lem:composition}]\quad\\ 
With some extra work, Theorem \ref{thm:severel:relations} implies the existence of 
Lagrangian relations for the data $\cC_1,\ldots,\cC_N\in \cP_\Delta$ 
with $\cC_{k+1}\neq \cC_k$. This follows, since the restriction of the functional 
\eqref{I:functional} to the open dense subset
$\Xi_{\cC_{N-1}} \times \ldots \times \Xi_{\cC_2}$ is not only smooth and
strictly convex, but also has been shown to have strictly positive second derivative.
So by the implicit function theorem the minimisers depend smoothly on  
$q_1\in \Xi_{\cC_1}$ and $q_N\in \Xi_{\cC_N}$.\\
The proof of Lemma \ref{lem:composition} is much more involved than the one of 
Theorem \ref{thm:severel:relations}. Moreover, it treats only the case 
$N=3$ (the composition of two Lagrangian relations).
In that case and for
$(x_1,x_3) \in L_{\cC_2,\cC_3;E_2,E_3} \circ L_{\cC_1,\cC_2;E_1,E_2}$
with $p_3^I=0$, one can uniquely reconstruct $x_2$ as well as the
$q_i\in \Delta_{\cC_i}$ ($i=1,3$), and then $q_2$ is the unique minimiser of $I_{q_3,q_1}$. 
\hfill $\Diamond$
\end{remark}

\addcontentsline{toc}{section}{References}
\end{document}